# Deep uncertainties in sea-level rise and storm surge projections: Implications for coastal flood risk management

Perry C. Oddo[1*], Ben S. Lee[2], Gregory G. Garner[3], Vivek Srikrishnan[4], Patrick M. Reed[5], Chris E. Forest[1,3,6,] and Klaus Keller[1,3,7]

[1]Department of Geosciences, The Pennsylvania State University, University Park, Pennsylvania, USA
[2]Department of Statistics, The Pennsylvania State University, University Park, Pennsylvania, USA
[3]Earth and Environmental Systems Institute, The Pennsylvania State University, University Park, Pennsylvania, USA
[4]Department of Energy and Mineral Engineering, The Pennsylvania State University, University Park, Pennsylvania, USA
[5]School of Civil and Environmental Engineering, Cornell University, Ithaca, New York, USA
[6]Department of Meteorology, The Pennsylvania State University, University Park, Pennsylvania, USA
[7]Department of Engineering and Public Policy, Carnegie Mellon University, Pittsburgh, Pennsylvania, USA

* Corresponding author. E-mail: pcoddo@gmail.com

**ABSTRACT**

Sea-levels are rising in many areas around the world, posing risks to coastal communities and infrastructures. Strategies for managing these flood risks present decision challenges that require a combination of geophysical, economic, and infrastructure models. Previous studies have broken important new ground on the considerable tensions between the costs of upgrading infrastructure and the damages that could result from extreme flood events. However, many risk-based adaptation strategies remain silent on certain potentially important uncertainties, as well as the trade-offs between competing objectives. Here, we implement and improve on a classic decision-analytical model (van Dantzig 1956) to: (i) capture trade-offs across conflicting stakeholder objectives, (ii) demonstrate the consequences of structural uncertainties in the sea-level rise and storm surge models, and (iii) identify the parametric uncertainties that most strongly influence each objective using global sensitivity analysis. We find that the flood adaptation model produces potentially myopic solutions when formulated using traditional mean-centric decision theory. Moving from a single-objective problem formulation to one with multi-objective trade-offs dramatically expands the decision space, and highlights the need for compromise solutions to address stakeholder preferences. We find deep structural uncertainties that have large effects on the model outcome, with the storm surge parameters accounting for the greatest impacts. Global sensitivity analysis effectively identifies important parameter interactions that local methods overlook, and which could have critical implications for flood adaptation strategies.

# 1. INTRODUCTION

Anthropogenic climate change is causing global sea-level to rise.[1–4] Projections of future sea-level rise are so uncertain that experts cannot agree on their associated probability distributions, a condition known as deep uncertainty.[5–9] For example, some recent estimates for potential upper bounds of global sea-level rise vary by more than one meter by the year 2100.[2,10–16] Coastal storm surge patterns are also deeply uncertain, and climate change has the potential to produce important nonstationarities in the most extreme storm events.[17,18] These deep uncertainties have critical implications for flood risk management, as even relatively small sea-level changes can increase flood risks by several orders of magnitude.[4] The combined effects of sea-level rise, land subsidence, changes in storm surge intensity, and economic growth pose immense threats to many coastal and low-lying communities.[19–23]

Strategies to manage flood risks are often analyzed with the help of integrated models that combine engineering, economic, and geophysical factors.[24–26] Following the devastating 1953 North Sea Flood, the Dutch government commissioned a study to determine coastal protection standards capable of withstanding future flood events. The resulting analysis, van Dantzig (1956), used a cost-benefit framework that balanced the costs of dike construction against the expected damages from flooding.[27] By minimizing the net present value of the project's total costs, the model determined optimal design levels corresponding to exceedance frequencies between 1/1,250 and 1/10,000 per year.[28,29]

This cost-benefit framework is still widely used in flood risk management applications.[30–34] Subsequent studies have expanded on the state-of-the-art using the van Dantzig (1956) analysis as a case study. Brekelmans et al.,[35] for example, introduces the problem of non-homogenous dike rings with independent heightenings and time horizons.

Further studies have included investigations into the economic benefits of learning as new information about future water levels is acquired.[36] Although these studies have broken important new ground, they still can be further improved in several aspects. For one, the original analysis uses a single-objective problem formulation (i.e., to minimize the discounted total costs over the lifetime of the project). By focusing on a single economic performance metric *a priori*, decision-makers may inadvertently restrict themselves to a limited set of possible solutions, a condition known as "cognitive myopia."[37–39] Giuliani et al.[40] define myopia in this context as a strategy that "fail[s] to explore the full set of trade-offs between evolving […] objectives and preferences," which contrasts other definitions which describe myopia as considering a limited time horizon.[41] Myopic decision-making can often result in degraded performance, especially when compared to multi-objective problem formulations which consider potentially conflicting preferences of a diverse set of stakeholders.[42,43]

The second area of potential improvement is to expand the consideration of model uncertainties. The parameter values used in van Dantzig (1956) were deterministic choices that reduced complex economic and geophysical mechanisms to best-guess estimates (see section 2.2. for a discussion of parameter value choices). Additionally, the original analysis makes several structural assumptions, particularly with respect to sea-level rise and storm surge projections. Model structure uncertainties can have large impacts on the expected risk of flooding.[44] Deconto and Pollard,[16] for example, recently demonstrated how accounting for previously underestimated ice-stability mechanisms can almost double previous estimates of Antarctic sea-level rise.

More modern flood risk models can support a sophisticated combination of digital topography, synthetic storm tracks, and multiple failure mechanisms to simulate coastal flood scenarios.[45,46] While simple models like van Dantzig (1956) are unable to achieve the spatial or temporal resolution available in more comprehensive studies, they are nevertheless useful for illustrating the impacts of uncertainties on flood adaptation in a clear and transparent way. Here we expand on the current state-of-the art by using the van Dantzig (1956) analysis as a didactic example to address three main questions:

1. What are the multi-objective trade-offs between key management objectives?
2. What are the effects of structural and parametric uncertainties in a classic economic model for flood adaptation?
3. Which parametric uncertainties matter the most for a given objective?

The following sections provide an overview of the model and outline the experimental methods, results, and research caveats. We first expand upon the single-objective formulation to include three additional management objectives and identify areas where competing preferences cause tension. We then describe the van Dantzig (1956) baseline model structure, and evaluate three versions of the analysis with increasing representational completeness. We accomplish this by introducing (i) parametric uncertainties, (ii) an improved sea-level rise model, and (iii) an improved storm surge model. Finally, we use global sensitivity analysis to determine which parameter uncertainties have the greatest impacts on the model outputs and objectives.

## 2. VAN DANTZIG (1956) MODEL OVERVIEW

### 2.1. Model description

The van Dantzig (1956) analysis was designed to determine optimal levels of flood protection for an individual dike ring area, given a set of geophysical and economic parameter inputs. The model uses a single decision variable: the height to which to raise a dike or levee.

Increasing dike heights lead to higher infrastructure expenditures, but result in decreased flood probabilities. Investment costs, *I*, are approximated by a linear increase with dike height:

$$I = kX, \quad (1)$$

where *X* is the desired dike height (in meters) and *k* is the cost of heightening by one meter (see Table I for a full list of parameter symbols, values, and units). Investment costs are fixed for the duration of the project's time horizon, so that no adaptive heightening can occur until the next decision is made to heighten.

Flood damages are determined by calculating the transient flood frequency for each year of the infrastructure's 75-year lifespan. The model assumes a stationary initial flood frequency for a given starting dike height, and evaluates exceedance probabilities for each considered dike height increase. The expected damages, *L*, are a product of the transient flood frequency and the discounted value of goods being protected:

$$L = p_0 e^{(-\alpha H_E)} V \frac{1}{(1+\delta')^t}, \quad (2)$$

where $p_0$ is the initial flood frequency, $\alpha$ is the exponential flood frequency decay rate, *V* is the value of goods contained in the dike ring area, $H_E$ is the effective dike height (i.e., the current height minus the effects of sea-level rise and land subsidence), and $\frac{1}{(1+\delta')^t}$ is the discount factor. The effective discount rate ($\delta'$), the rate of sea-level rise ($\phi$), and the rate of land subsidence ($\eta$) are all assumed to be constant for every year, *t*.

The expected flood damages are discounted across the project's time horizon and balanced against the costs of investing into infrastructure. Investment costs and expected damages are evaluated for each considered dike height. Both curves are then summed to produce a curve for total costs. The optimal dike heightening for this model formulation is the minimum along the total costs curve (Fig. S1).

**2.2. Parameter Value Selection**

The baseline model version uses the parameter values outlined in section 6 ("The Doubtful Constants") of van Dantzig (1956). The values themselves are estimates designed to approximate the geophysical and economic dynamics of the original site in Hoek van Holland, the Netherlands. Note that choosing these parameter values from the van Dantzig (1956) study requires some interpretation.

The value of goods, *V,* for example, attempts to account for cost of human life by including a "more or less [arbitrary]" factor of two, instead of explicitly assigning an economic valuation to each life lost.[27] Subsequent studies have attempted to assign a tangible cost to flood victims, yet the results are highly variable. Vrijling et al.[47] suggest using the present value of the per capita Net National Product (NNP), which frames loss of life in the context of the national GDP. Other cost frameworks incorporate macro-economic indicators like expected lifespan or Life Quality Index (LQI) to produce estimates ranging from 500,000 to 4 million Euros per life.[48] For the purposes of this didactic approach, we retain the original factor of two, but note its limited efficacy for modern decision-making frameworks.

**3. METHODS**

This study builds upon the van Dantzig (1956) analysis in several ways. We first introduce additional explicit management objectives to identify the multi-objective tradeoffs common to more complex decision problems (section 3.1 – Multi-objective Trade-off Analysis). Second, we investigate the effects of model structural uncertainties by contrasting the baseline model with three increasingly complex model formulations (section 3.2 – Model Development).

Finally, we perform sensitivity analyses to identify the parameters most responsible for influencing model output (section 3.3 – Sensitivity Analysis). The following sections details the methods we use for each component.

**3.1 Multi-objective Trade-off Analysis**

Identifying key management objectives in a decision framework is a critical step toward accounting for stakeholders' preferences and avoiding potentially myopic biases.[8,38,49] The sole objective in the van Dantzig (1956) problem formulation minimizes the net present value of total flood protection costs ($O_1$). This problem formulation neglects all uncertainties and assumes perfect foresight when arriving at its optimal solution. Systems optimized in this classical sense can be very sensitive to small perturbations from assumed conditions.[50,51] Moreover, flood management is a classical archetype of "wicked" social planning problems.[52] These problems typically possess a broad array of conflicting objectives and severe uncertainties. In this study, we demonstrate the value of moving beyond the tacit assumptions and tradeoffs intrinsic to the deterministic single-objective formulation, to a more explicit tradeoff analysis across performance and cost objectives. We therefore consider three additional management objectives that are readily derived from the existing economic framework:

1. Minimize the costs of investing into dike construction ($O_2$): The objective function is,

$$Investment\ costs = kX_i, \tag{3}$$

    where, $X$ is the considered dike height for each height index, $i$, and $k$ is the cost rate of heightening by one meter.

2. Minimize probability of flooding ($O_3$): This objective aims to minimize the expected likelihood of the current dike being overtopped for each year of the project's time horizon (T=75). The objective function is,

$$Flood\ probability = E\left[\frac{1}{T}\sum_{t=1}^{T} p_0 e^{(-\alpha H_{E_{i,t}})}\right]_N \quad (4)$$

where, $p_0 e^{(-\alpha H_{E_{i,t}})}$ is the transient flood frequency for each effective dike height, $H_{Ei}$, at each time step, $t$ and height index, $i$. The notation $E[\ ]_N$ denotes the expectation over the $N^{th}$ State of the World (SOW).

3. Minimize the discounted damages of flood events ($O_4$): This objective aims to minimize the expected damages to the discounted material goods protected by the dike ring. The objective function is,

$$Discounted\ damages = E\left[\frac{1}{T}\sum_{t=1}^{T} p_0 e^{(-\alpha H_{E_{i,t}})} V \frac{1}{(1+\delta')^t}\right]_N. \quad (5)$$

### 3.2. Model Development

*3.2.1 Model Version #1: Baseline*

The baseline version of the model was transcribed from van Dantzig (1956) using the best-guess parameter estimates provided in the original paper. We evaluate our four objectives for each dike height from zero to ten meters in five-centimeter increments. The results of our model and parameter choices reproduce the optimal height described by van Dantzig (1956) within a 5% margin of error (Fig. S1).

*3.2.2. Model Version #2: Parametric Uncertainties*

For the second model version, we introduce parametric uncertainties into the baseline model while retaining the same basic structure. To accomplish this, we generate an ensemble of 10,000 possible parameter values using random Latin Hypercube samples from prior marginal distributions (Fig. S2).[53] Where possible, we derived marginal parameter distributions using published estimates from the literature (e.g. land subsidence rate,[54] cost rate of heightening,[55,56] and effective discount rate).[57] Where published estimates were not available, we assume prior distributions are normal or lognormal (based on whether or not each parameter can physically fall below zero), and use the baseline parameter values as distribution means (Table I). Each of the 10,000 parameter sets represents an uncertain State of the World (SOW). In addition to enumerating how our objectives perform for each parameter set, we also identify the expected (mean) performance across all SOW.

*3.2.3. Model Version #3: Improved Sea-Level Rise Analysis*

Sea-level projections in van Dantzig (1956) are assumed to increase at a constant annual rate. For the upgraded sea-level rise model, we adopt the framework in Lempert et al.,[26] which approximates mean annual sea-level projections, *SLR*, as

$$SLR = a + bt + ct^2 + c^*I(t - t^*). \qquad (6)$$

In this model, the parameters *a*, *b*, and *c* represent the reasonably well-characterized process of thermosteric expansion as a 2$^{nd}$-order polynomial.[26,58] It also accounts for more poorly-understood processes, including potential abrupt sea-level rise consistent with sudden changes in ice flow dynamics.[59] Here, $c^*$ represents an increase in the rate of sea-level rise which takes place at some uncertain time, $t^*$, in the future.

We calibrate the sea-level model using tide gauge observations from Delfzijl, the Netherlands (Fig. S3). We fit mean annual tide gauge observations from a 137-year record to a 2nd-order polynomial approximation using a linear regression model. We then bootstrapped the residuals of the polynomial fit to produce a time-series of 55,000 sea-level hindcasts with approximately the same autocorrelation structure as the original tide gauge observations. Projections for sea-level at the year 2100 were made using the model described in equation 6 (Fig. 1). The full suite of projections varied according to uncertainties in the five sea-level rise parameters, with the timing and rate increase parameters (c* and t*, respectively) causing the majority of the variation. We use the probabilistic inversion method described in Lempert et al.[26] to constrain the sea-level projections so that they are consistent with an expert assessment of sea-level rise in the year 2100 (represented here by a modified beta distribution). This method uses rejection sampling to selectively remove future sea-level projections which fall outside the bounds of the beta distribution expert assessment. The result of the rejection sampling is a joint probability distribution for each of the five sea-level rise parameters that can then be run through the model (Fig. S4).

*3.2.4. Model version #4: Improved Storm Surge Analysis*

The van Dantzig (1956) analysis estimates storm surge rates by plotting the exceedance frequencies from 49 years of maximum annual tide gauge observations on a semi-log plot. The exceedance frequencies are well approximated by a linear trend with no clearly apparent tendency to vertically asymptote at higher return levels. This method produces a reasonable fit to the data (see Fig. S5), yet van Dantzig (1956) notes that it is arguably a poor assumption when projecting storm surges for very long return periods.

Extreme storm surge events have disproportionately large societal impacts and require large quantities of data to project accurately.[60–62] In the United States, the Federal Emergency Management Agency (FEMA) recommends using a Generalized Extreme Value (GEV) technique where extended observational records (more than 30 years) are available.[63,64] Here, we follow this methodology to extrapolate the data from an extended (137-year) tide gauge record from the Dutch Ministry of Infrastructure and the Environment.[65–67] The observation frequencies vary from three hours to ten-minute increments. We calculate the residuals from the annual mean sea level in order to remove long-term variability (Fig. S6). We then aggregate the detrended records into annual block maxima, and fit them to the GEV distribution using a maximum likelihood estimate method (MLE).[65] The GEV cumulative distribution is described by the equation

$$F(x;\mu,\sigma,\xi) = \exp\left\{-\left[1 + \xi\left(\frac{x-\mu}{\sigma}\right)\right]^{-\frac{1}{\xi}}\right\}, \tag{7}$$

where μ is the location parameter, σ is the scale parameter, and ξ is the shape parameter. We use the resulting MLE estimates for the three GEV parameters to determine flood return levels with uncertainties up to the 1/10,000 return level.[68]

To provide some insight into the effects of methodological uncertainty, we also contrast this frequentist approach with a Bayesian approach. We employ Markov Chain Monte Carlo (MCMC), specifically the Metropolis-Hastings (MH) algorithm, to draw samples from the posterior distribution of the three GEV parameters. Each parameter's prior distribution is a normal distribution centered at that parameter's maximum likelihood estimate. We run the MH algorithm for 100,000 iterations, and trace plots of the MCMC chains are consistent with the hypothesis that the chains converged. We use the resulting samples from the parameters' posterior distribution to (1) generate samples of the return levels for a given return period; and (2) measure the uncertainty underlying the parameters and return levels[69–72]. For the return level samples, the posterior distribution sample mean represents the expected return level for a given return period, and the highest posterior density (HPD) intervals were calculated to determine both the 90% and 95% credible intervals (Fig. 2; Fig. S7).[73] Reis and Stedinger[70] note that this comparatively more comprehensive method produces more accurate approximations of credible intervals for flood quantiles than the frequentist MLE estimations discussed above.

### 3.3 Sensitivity Analysis

*3.3.1. One-at-a-time (OAT) Local Sensitivity Analysis*

Sensitivity analyses are important for determining how variations in a model's input can produce uncertainty in the model's output.[74,75] They serve as diagnostic tools to isolate the parameters that may require additional calibration and to identify potential knowledge gaps.[74,76] A one-at-a-time (OAT) sensitivity analysis is a simple method useful for characterizing linear systems. It is known as a local method because it quantifies the extent to which an individual parameter affects model output.[77,78] We apply a OAT analysis by varying each single parameter from the 1$^{st}$–99$^{th}$ percentile of its prior distribution while holding all others constant. The results of the van Dantzig (1956) model then serve to rank the parameters in order of their impact on the model variance.[8,75,78]

*3.3.2. Sobol' Method Global Sensitivity Analysis*

While this OAT analysis is a useful metric to infer individual parameter effects, it is unable to identify parameter interactions. Local methods only evaluate parameter sensitivities for a single choice of the other parameters, instead of the full parameter space.[79] Global methods, in contrast, determine sensitivities by varying all parameters simultaneously.[74,80] We therefore also apply the global variance-based Sobol' method to illustrate how potentially important parameter interactions may go unnoticed when relying on local methods.[74,78] Sobol' sensitivity analyses traditionally use uniform parameter distributions when performing variance decomposition. To fully utilize our non-uniform distributions, we first sampled each parameter uniformly between zero and one to produce a set of sampling indices. We then use the resulting values to sample percentiles over the inverse cumulative distribution functions (CDF) for our desired distributions. The analysis returns a series of Sobol' indices, which characterize the impacts of individual parameters on the model output, in addition to identifying parameter interaction terms.

## 4. RESULTS AND DISCUSSION

The original van Dantzig (1956) analysis represents a relatively simple treatment of a nontrivial decision problem. Here, we attempt to discuss how incorporating deeply uncertain geophysical processes and improved statistical inferences can rapidly increase the complexity of an otherwise didactic problem. We discuss the effects of these uncertainties by considering how each subsequent analytical improvement impacts the model output.

### 4.1. Uncertainty Quantification

*4.1.1. Parametric Uncertainties*

The simple structure of the original van Dantzig (1956) analysis relied on perfect foresight in its treatment of model uncertainties. Using the single-objective problem formulation, the baseline version of the model determines an optimal dike heightening of 2.35 meters (Fig. 3A). Adding parametric uncertainties to the model imparts considerable variability for the investment costs and discounted damages (Fig. 3B). The gray line in the parametric uncertainty plot shows the curve for the expected total costs, averaged across all 10,000 SOW. We focus on the expected (average) outcome due to its emphasis in classic decision theory, which states that a rational agent seeks to optimize expected utility in an uncertain world.[81] In doing so, we can compare the minimum of the expected total costs with the baseline model scenario. We note that introducing parametric uncertainties has a relatively small effect on the new 'optimal' dike heightening (five centimeter increase), suggesting the model structure is fairly linear.

*4.1.2. Structural Uncertainties*

Changes in structural model uncertainties (i.e., sea-level rise and storm surge determinations) have comparatively larger effects on the model outcomes. The van Dantzig (1956) linear sea-level rate (8 mm/year) projects a mean sea-level estimate of ~0.7 meters by the year 2100 (green line, Fig. 1). This estimate is lower than approximately 85% of the beta-calibrated forecasts, which project a mean sea-level of ~1.2 meters. Including the updated sea-level rise model into the van Dantzig (1956) formulation increases the variability in the expected model damages (width of red envelop across all SOW, Fig. 3C). We see an approximately 11% increase in the expected 'optimal' total cost index, from 2.35 to 2.6 meters. It should be noted that this 0.25-meter increase may appear rather small, but across the entire ring area it corresponds to an investment cost of approximately 11 million guilders.

Refining the storm surge model using generalized extreme value analysis has much more apparent effects on the projected outcomes. Extreme storm events have the potential to contribute multiple meters of additional storm surge, contrasted with the smaller contribution from the sea-level-rise model. Even accounting for abrupt sea-level dynamics, only about 95 centimeters of expected sea-level is projected by the end of the 75-year project time horizon. On longer time scales, the van Dantzig (1956) storm surge approximation (i.e. linear fit through the annual maximum return levels) produces an extrapolation of around eight meters for the 1/10,000 return period. This projection is higher than the expected value of the Bayesian MCMC method (6.78 meters), yet still within the 95% and 90% credible intervals (Fig. 2).

We also note that the Bayesian MCMC projects higher expected return levels for each considered return period than the frequentist MLE method (Fig. S7). The upper bounds of the MCMC 90% confidence intervals are also consistently higher than when using the MLE approach. In the United States, the U.S. Army Corps of Engineers uses this upper 90% confidence interval when evaluating levee systems under the National Flood Insurance Program.[82] Specifically, to satisfy evaluation requirements "…a levee or incised channel must have at least a 90% assurance of excluding the 1% annual chance exceedance flood for all reaches of the system," (i.e., the 100-year flood).[83] Constraining the upper 90% confidence interval through methodological or statistical improvements could therefore have important implications for flood adaptation policy.

When incorporated into the van Dantzig (1956) formulation, the upgraded storm surge model increases the expected 'optimal' dike heightening by roughly 91% to 4.5 meters (Fig. 3D). This marked increase is driven by the large variability in the expected damages caused by extreme storm surge events. Both structural model improvements—but particularly the upgraded storm surge model—illustrate the potential bias inherent in mean-centric decision analyses. The 'optimal' heightening determined here may satisfy the expected outcome according to the original single-objective problem formulation. However, it remains sensitive to the system's significant uncertainties, and would likely be subject to flooding due to the deeply uncertain storm surge patterns.[50] Nearly 20% of the model permutations produce results higher than the 4.5 meter expected outcome, each of which represents a possible SOW in which significant damages could have been incurred.

**4.2. Multi-objective Trade-offs**

The investment cost objective produces clear trade-offs when plotted against the other objectives (Fig. 4). Here, we can see the expansion of the objective space relative to (i) the deterministic baseline model (dashed line) and (ii) the expected trade-off across all SOW (solid line). The black points show the minimum values along those respective trade-off curves. The black star in the lower left corner of each plot represents the hypothetical ideal point. In a multi-objective framework, the ideal point may not actually be feasible, but instead illustrates the best values that are attainable for each objective considered by itself (for example, zero investment costs for flood protection while incurring zero flood damages).[38]

The baseline van Dantzig (1956) tradeoff results (Fig. 4) illustrate the myopia of the original deterministic baseline solution. The dashed trade-off curves represent the set of perfect foresight dike heightening decisions that compose the deterministic Pareto front (i.e., those solutions whose performance in one objective cannot be improved with degrading the performance in one or more the remaining objectives).[38,43,84] With the addition of parametric and structural uncertainties, we see the translation of the mean performance tradeoffs as would be expected from the perfect foresight deterministic case. The solid curves represent the solutions determined using a traditional mean-centric decision analysis approach. By adhering to the principle of rational, expected behavior, we see both the Pareto front and the 'optimal' solution shift further from the ideal point (i.e., the multi-objective effects of imperfect knowledge). Moreover, the background color contours illustrating the relative densities of potential SOW emphasize that mean-centric decision analysis using only the three mean solid curve tradeoffs (Fig. 4) could have very severe and potentially irreversible consequences. For example, the linearly scaled tradeoffs in the top panels show that seeking to minimize expected investment costs yields a rapid growth in the variance of the potential realized performance in either discounted damages or total costs. More concerning, the log-scaled bottom panel shows that decisions based on the solid mean tradeoff curve yields significant residual risks that the realized system protection is very far from the intended level in the tradeoff analysis. The minimum expected NPV of total costs solution is intended to provide the 1/10,000 year design level but has a strong potential to actually provide less than the 1/100 year design level of protection.

It is important to explore the decision relevance of the residual risks implicit to choices made based on the expected value tradeoffs (Fig. 4). For example, one can define what potential realized performance levels are acceptable (or "satisficing") through setting multivariate thresholds (Fig. 5).[8] In this illustration, hypothetical decision-makers can set objective thresholds under which they prefer to stay. The first would be to maintain an expected annual flood probability of under 1/10,000 per year, and the second would be to invest less than 100 million guilders for protection costs (Panel A, blue box). Of the more than 2 million solutions evaluated between these two objectives, only 1.1% meet both criteria. The dashed line again shows how the van Dantzig (1956) baseline solution would have performed under these threshold constraints. The expanded view shows a situation in which a third threshold is instituted, such as restricting flood damages to less than 1 million Guilders (Panel B). In this scenario, only the solutions highlighted in green full satisfy all three threshold requirements. The scarcity of acceptable solutions—in this case, only about 0.20% of all possible solutions—illustrates the need to understand how model uncertainties can map onto the objective space in a decision-making framework.

**4.3. Sensitivity Analysis**

The diagnosed parameter sensitivities vary greatly depending on whether local or global methods are used, and depending on which version of the analysis is being considered. One-at-a-time (OAT) sensitivities for the upgraded sea-level rise model, for example, show that the parameters governing abrupt sea-level rise—the timing, $t^*$, and rate, $c^*$—have a strong influence over three of our objectives (total costs, discounted damages, and flood probability, Fig. S8). The width of the bar next to the parameter description indicates the magnitude of the sensitivity directly attributable to that parameter.[77,78] When the OAT methods are applied to the upgraded storm surge model, however, the GEV parameters dominate the sensitivities instead (Fig. 6). This is consistent with the observation discussed above that large storm surge events can overpower the signal of the comparatively smaller contributions from sea-level rise. Intuitively, for every model version the cost rate of heightening parameter, $k$, accounted for all of the variance for the investment costs objective (see equation 3).

The global sensitivity analysis produces similar rankings driven by storm surge and economic parameters (Fig. 7). The size of the points at each node indicates the magnitude of the first and total-order effects from the variance decomposition analysis.[78] First-order effects illustrate the effects of the individual parameter on the model output while total-order effects are a measure of overall parameter weight (individual plus interaction terms). Lines of increasing thickness represent the magnitudes of second-order interactions. Here too, the investment cost objective is determined solely by the $k$ value. The GEV parameters are important drivers of discounted total costs, flood probability, and discounted damages, with the largest share (between ~77% and ~84%) attributable to the shape parameter ($\xi$). Even where large first-order sensitivities don't exist, interactions between the GEV parameters produce sizeable total-order sensitivities (e.g., scale parameter in Panel A). Such interaction terms are critical to understanding the model behavior, as the shape and scale parameters, in conjunction, most directly influence how the extreme tail-area values are extrapolated.[66,85] The effects of the sea-level rise parameters are overshadowed in the most upgraded version of the model, likely driven to a large extent by the fact that sea-level projections are only considered over a 75-year investment period. Over longer time horizons, we hypothesize that the effects of abrupt sea-level rise to register a larger impact in the sensitivity analyses (see the large second-order interactions in the upgraded sea-level rise model, Fig. S9).

## 5. CAVEATS AND FUTURE RESEARCH NEEDS

We have adopted the economic decision analysis described in van Dantzig (1956) as a didactic example of how to apply a relatively simple decision framework to a complex problem. Using a simple model helps to keep the analysis relatively transparent and concise. However, this simplicity also points to potentially important caveats and future research needs.

Future climate change projections are deeply uncertain, as are their likely impacts on coastal adaptation frameworks. Our analysis introduces updated models for both sea-level rise and storm surge, yet remains silent on a number of additional uncertainties. As previously mentioned, several of the parameter values used in the original analysis could be further constrained. For example, our analysis assumes a simple stationary discount rate of 2%, yet choosing an appropriate discount rate remains a contentious issue in the climate literature.[86–90] Furthermore, the cost rate of heightening, $k$, is approximated as constant. In reality, heightening a dike would necessarily require widening the base. The additional material required, combined with the potential for land reclamation may result in a cost function more closely resembling a quadratic or exponential shape.[35] For parameters that did not have published ranges available, we assume either normal or lognormal distributions using the baseline values as distribution means. This assumption was intended to produce illustrative ranges for this specific area of interest. As this assumption will likely have important implications for the sensitivity analyses, it is all the more critical to further constrain parameter distributions in future work, and to investigate model uncertainties using a deep uncertainty framework.

While this analysis included three additional management objectives, it still only considers a single decision variable (i.e. how high to build the dike). Future studies could incorporate numerous other decision variables, including varying the timing of dike heightening or investing into climate resilience or mitigation in addition to adaptation. Louisiana's 2017 Comprehensive Master Plan for a Sustainable Coast is one recent example of how to implement an integrated portfolio of flood management strategies.[91] Over a 50 year time horizon, this master plan strengthens structural protections while simultaneously incorporating marsh creation, sediment diversion, barrier island restoration, and other adaptive management measures. As additional variables are introduced, the considered decision space rapidly expands. Applying advanced analytical techniques like multi-objective evolutionary algorithms (MOEAs) would allow for decision-makers to efficiently explore trade-offs and solutions in increasingly complex problem formulations.[39,92]

## 6. CONCLUSION

This study reveals how a simple flood risk model (van Dantzig 1956) can be used to investigate the effects of methodological assumptions and model uncertainties on a decision-analytical framework. Our results demonstrate three important points. First, a simple implementation of the traditional decision theory can produce myopic solutions in flood adaptation frameworks, as can be seen from the changes associated with moving from single-objective problem formulations to diverse multi-objective tradeoffs. Seeking compromise solutions by satisficing some optimal performance for more stability is an effective technique for avoiding extreme perspectives and addressing multiple stakeholder preferences. Second, we find that accounting for model uncertainties—particularly the structural representations of sea-level-rise and storm surges—can result in large changes to the suggested economically optimal solution. Finally, we find that model sensitivities change dramatically depending on which model version is being considered and whether local or global sensitivity methods are employed. Sobol' Sensitivity Analysis effectively identified useful parameter interactions which were otherwise overlooked, and which could have critical implications for flood adaptation strategies.

**ACKNOWLEGEMENTS**

The authors would like to thank the members of the Keller, Forest, and Reed research groups for thoughtful discussions about this work. Special thanks to Alexander Bakker for particular insight into Dutch culture and flood management strategies, and to Vivek Srikrishnan for reviewing and reproducing the model code. This work was partially supported by the National Science Foundation (NSF) through the Network for Sustainable Climate Risk Management (SCRiM) under NSF cooperative agreement GEO-1240507 and the Penn State Center for Climate Risk Management. Any opinions, findings, and conclusions or recommendations expressed in this material are those of the author(s) and do not necessarily reflect the views of the National Science Foundation. All coauthors contributed to the interpretation of results, writing, and revision of the manuscript.

**ELECTRONIC SUPPLEMENTARY MATERIAL**

The code required to reproduce this analysis will be posted prior to publication at: https://github.com/

**REFERENCES**

1. Jevrejeva S, Moore JC, Grinsted A. How will sea level respond to changes in natural and anthropogenic forcings by 2100? Geophysical Research Letters, 2010; 37(7).

2. Church JA, Clark PU, Cazenave A, Gregory JM, Jevrejeva S, Levermann A, Merrifield MA, Milne GA, Nerem RS, Nunn PD, Payne AJ, Pfeffer WT, Stammer D, Unnikrishnan AS. 2013: Sea Level Change. Pp. 1137–1216 in Stocker TF, Qin D, Plattner G-K, Tignor M, Allen SK, Boschung J, Nauels A, Xia Y, Bex V, Midgley PM (eds). Climate Change 2013: The Physical Science Basis. Contribution of Working Group I to the Fifth Assessment Report of the Intergovernmental Panel on Climate Change. Cambridge, United Kingdom: Cambridge University Press, 2014.

3. Strauss BH, Kulp S, Levermann A. Carbon choices determine US cities committed to futures below sea level. Proceedings of the National Academy of Sciences, 2015:201511186.

4. Sriver RL, Urban NM, Olson R, Keller K. Toward a physically plausible upper bound of sea-level rise projections. Climatic Change, 2012; 115(3–4):893–902.

5. Keller K, Nicholas R. Improving climate projections to better inform climate risk management. Pp. 9–18 in Bernard L, Semmler W (eds). The Oxford Handbook of the Macroeconomics of Global Warming. Oxford University Press, 2015.

6. Walker WE, Lempert RJ, Kwakkel JH. Deep uncertainty. Pp. 395–402 in Encyclopedia of Operations Research and Management Science. Springer, 2013.

7. Knight FH. Risk, uncertainty and profit. Hart, Schaffner and Marx, 1921.

8. Herman JD, Reed PM, Zeff HB, Characklis GW. How Should Robustness Be Defined for Water Systems Planning under Change? Journal of Water Resources Planning and Management, 2015:04015012.

9. Kwakkel JH, Walker WE, Marchau V. From predictive modeling to exploratory modeling: how to use non-predictive models for decisionmaking under deep uncertainty. P. in Proceedings of the 25th Mini-EURO Conference on Uncertainty and Robustness in Planning and Decision Making (URPDM2010), 15-17 April. 2010.

10. Jevrejeva S, Grinsted A, Moore JC. Upper limit for sea level projections by 2100. Environmental Research Letters, 2014; 9(10):104008.

11. Horton BP, Rahmstorf S, Engelhart SE, Kemp AC. Expert assessment of sea-level rise by AD 2100 and AD 2300. Quaternary Science Reviews, 2014; 84:1–6.

12. Mengel M, Levermann A, Frieler K, Robinson A, Marzeion B, Winkelmann R. Future sea level rise constrained by observations and long-term commitment. Proceedings of the National Academy of Sciences, 2016; 113(10):2597–2602.

13. Kopp RE, Horton RM, Little CM, Mitrovica JX, Oppenheimer M, Rasmussen DJ, Strauss BH, Tebaldi C. Probabilistic 21st and 22nd century sea-level projections at a global network of tide-gauge sites. Earth's Future, 2014; 2(8):383–406.

14. Katsman CA, Sterl A, Beersma JJ, Brink HW van den, Church JA, Hazeleger W, Kopp RE, Kroon D, Kwadijk J, Lammersen R, Lowe J, Oppenheimer M, Plag H-P, Ridley J, Storch H von, et al. Exploring high-end scenarios for local sea level rise to develop flood protection strategies for a low-lying delta—the Netherlands as an example. Climatic Change, 2011; 109(3–4):617–645.

15. Parris A, Bromirski P, Burkett V, Cayan D, Culver M, Hall J, Horton R, Knuuti K, Moss R, Obeysekera J, Sallenger A, Weiss J. Global Sea Level Rise Scenarios for the United States National Climate Assessment. Climate Program Office, Silver Springs, MD: National Oceanic and Atmospheric Administration, 2012:37.

16. DeConto RM, Pollard D. Contribution of Antarctica to past and future sea-level rise. Nature, 2016; 531(7596):591–597.

17. Grinsted A, Moore JC, Jevrejeva S. Projected Atlantic hurricane surge threat from rising temperatures. Proceedings of the National Academy of Sciences, 2013; 110(14):5369–5373.

18. Rosner A, Vogel RM, Kirshen PH. A risk-based approach to flood management decisions in a nonstationary world. Water Resources Research, 2014; 50(3):1928–1942.

19. de Moel H, Aerts JCJH, Koomen E. Development of flood exposure in the Netherlands during the 20th and 21st century. Global Environmental Change, 2011; 21(2):620–627.

20. Maloney MC, Preston BL. A geospatial dataset for U.S. hurricane storm surge and sea-level rise vulnerability: Development and case study applications. Climate Risk Management, 2014; 2:26–41.

21. Hanson S, Nicholls R, Ranger N, Hallegatte S, Corfee-Morlot J, Herweijer C, Chateau J. A global ranking of port cities with high exposure to climate extremes. Climatic Change, 2011; 104(1):89–111.

22. Kane HH, Fletcher CH, Frazer LN, Anderson TR, Barbee MM. Modeling sea-level rise vulnerability of coastal environments using ranked management concerns. Climatic Change, 2015; 131(2):349–361.

23. Hallegatte S, Green C, Nicholls RJ, Corfee-Morlot J. Future flood losses in major coastal cities. Nature climate change, 2013; 3(9):802–806.

24. Vrijling JK. Probabilistic design of water defense systems in The Netherlands. Reliability Engineering & System Safety, 2001; 74(3):337–344.

25. Doorn N. Rationality in flood risk management: the limitations of probabilistic risk assessment in the design and selection of flood protection strategies. Journal of Flood Risk Management, 2014; 7(3):230–238.

26. Lempert R, Sriver RL, Keller K. Characterizing Uncertain Sea Level Rise Projections to Support Investment Decisions. California Energy Commission Sacramento, CA, USA, 2012.

27. van Dantzig D. Economic Decision Problems for Flood Prevention. Econometrica, 1956; 24(3):276–287.

28. Jonkman SN, Bočkarjova M, Kok M, Bernardini P. Integrated hydrodynamic and economic modelling of flood damage in the Netherlands. Ecological Economics, 2008; 66(1):77–90.

29. Eijgenraam C, Kind J, Bak C, Brekelmans R, den Hertog D, Duits M, Roos K, Vermeer P, Kuijken W. Economically Efficient Standards to Protect the Netherlands Against Flooding. Interfaces, 2014; 44(1):7–21.

30. Dijkman J. A Dutch Perspective on Coastal Louisiana Flood Risk Reduction and Landscape Stabilization. Report WL-Z4307: U.S. Army Engineer Research and Development Center, 2007.

31. Kind JM. Economically efficient flood protection standards for the Netherlands. Journal of Flood Risk Management, 2014; 7(2):103–117.

32. Lund JR. Integrating social and physical sciences in water management. Water Resources Research, 2015; 51(8):5905–5918.

33. Mai CV, van Gelder PHAJM, Vrijling JK, Mai TC. Risk Analyis of Coastal Flood Defences - A Vietnam Case. Institute for Catastrophic Loss Reduction, 2008.

34. Hallegatte S. Uncertainties in the Cost-Benefit Analysis of Adaptation Measures, and Consequences for Decision Making. Pp. 169–192 in Linkov I, Bridges TS (eds). Climate. Dordrecht: Springer Netherlands, 2011.

35. Brekelmans R, Hertog D den, Roos K, Eijgenraam C. Safe Dike Heights at Minimal Costs: The Nonhomogeneous Case. Operations Research, 2012; 60(6):1342–1355.

36. van der Pol TD, van Ierland EC, Weikard H-P. Optimal dike investments under uncertainty and learning about increasing water levels: Optimal dike investments. Journal of Flood Risk Management, 2014; 7(4):308–318.

37. Hogarth RM. Beyond discrete biases: Functional and dysfunctional aspects of judgmental heuristics. Psychological Bulletin, 1981; 90(2):197–217.

38. Singh R, Reed PM, Keller K. Many-objective robust decision making for managing an ecosystem with a deeply uncertain threshold response. Ecology and Society, 2015; 20(3):12.

39. Kasprzyk JR, Nataraj S, Reed PM, Lempert RJ. Many objective robust decision making for complex environmental systems undergoing change. Environmental Modelling & Software, 2013; 42:55–71.

40. Giuliani M, Herman JD, Castelletti A, Reed P. Many-objective reservoir policy identification and refinement to reduce policy inertia and myopia in water management. Water Resources Research, 2014; 50(4):3355–3377.

41. Kurz M. Myopic Decision Rules. Pp. 136–141 in Eatwell J, Milgate M, Newman P (eds). Utility and Probability. The New Palgrave. Palgrave Macmillan UK, 1990.

42. Kasprzyk JR, Reed PM, Characklis GW, Kirsch BR. Many-objective de Novo water supply portfolio planning under deep uncertainty. Environmental Modelling & Software, 2012; 34:87–104.

43. Garner G, Reed P, Keller K. Climate risk management requires explicit representation of societal trade-offs. Climatic Change, 2016; 134(4):713–723.

44. McInerney D, Keller K. Economically optimal risk reduction strategies in the face of uncertain climate thresholds. Climatic Change, 2008; 91(1–2):29–41.

45. Fischbach JR, Louisiana, Rand Gulf States Policy Institute eds. Coastal Louisiana Risk Assessment Model: Technical Description and 2012 Coastal Master Plan Analysis Results. Santa Monica, Calif: Rand Corp, 2012.

46. Johnson DR, Fischbach JR, Ortiz DS. Estimating Surge-Based Flood Risk with the Coastal Louisiana Risk Assessment Model. Journal of Coastal Research, 2013:109–126.

47. Vrijling JK, van Hengel W, Houben RJ. A framework for risk evaluation. Journal of Hazardous Materials, 1995; 43(3):245–261.

48. Jonkman SN. Loss of life estimation in flood risk assessment; theory and applications. 2007. Ph.D. Dissertation, TU Delft, Delft University of Technology.

49. Brill ED, Flach JM, Hopkins LD, Ranjithan S. MGA: a decision support system for complex, incompletely defined problems. IEEE Transactions on Systems, Man, and Cybernetics, 1990; 20(4):745–757.

50. Beyer H-G, Sendhoff B. Robust optimization – A comprehensive survey. Computer Methods in Applied Mechanics and Engineering, 2007; 196(33–34):3190–3218.

51. Haasnoot M, Kwakkel JH, Walker WE, ter Maat J. Dynamic adaptive policy pathways: A method for crafting robust decisions for a deeply uncertain world. Global Environmental Change, 2013; 23(2):485–498.

52. Rittel HW, Webber MM. Dilemmas in a general theory of planning. Policy sciences, 1973; 4(2):155–169.

53. McKay MD, Beckman RJ, Conover WJ. A Comparison of Three Methods for Selecting Values of Input Variables in the Analysis of Output. Technometrics, 1979; 21(2):239–245.

54. Rietveld H. Land subsidence in the Netherlands. Pp. 455–465 in Proceedings of the 3rd International Symposium on Land Subsidence. Vol 151. 1986.

55. Aerts JCJH, Botzen WJW. Climate Adaptation Cost for Flood Risk Management in the Netherlands. Pp. 99–113 in Storm Surge Barriers to Protect New York City. American Society of Civil Engineers, 2013.

56. Hillen MM, Jonkman SN, Kanning W, Kok M, Geldenhuys MA, Stive MJF. Coastal Defence Cost Estimates: Case Study of the Netherlands, New Orleans and Vietnam. TU Delft, Department Hydraulic Engineering, 2010.

57. OECD, International Transport Forum. Discounting Long-term Effects of Climate Change for Transport. Pp. 53–72 in Adapting Transport Policy to Climate Change: Carbon Valuation, Risk and Uncertainty. Paris: OECD Publishing, 2015.

58. Church JA, White NJ. A 20th century acceleration in global sea-level rise. Geophysical Research Letters, 2006; 33(1):L01602.

59. Alley RB, Marotzke J, Nordhaus WD, Overpeck JT, Peteet DM, Pielke RA, Pierrehumbert RT, Rhines PB, Stocker TF, Talley LD, Wallace JM. Abrupt climate change. Science, 2003; 299(5615):2005–2010.

60. van den Brink HW, Kinnen GP, Opsteegh JD, van Oldenborgh GJ, Burgers G. Improving $10^4$-year surge level estimates using data of the ECMWF seasonal prediction system. Geophysical Research Letters, 2004; 31:L17210.

61. Cameron D. Flow, frequency, and uncertainty estimation for an extreme historical flood event in the Highlands of Scotland, UK. Hydrological Processes, 2007; 21(11):1460–1470.

62. Baart F, Bakker MAJ, van Dongeren A, den Heijer C, van Heteren S, Smit MWJ, van Koningsveld M, Pool A. Using 18th century storm-surge data from the Dutch Coast to improve the confidence in flood-risk estimates. Natural Hazards and Earth System Science, 2011; 11(10):2791–2801.

63. Wallace EE, MacArthur RC, Chowdhury S, Sakumoto LK. Coastal Flood Hazard Analysis and Mapping Guidelines for the Pacific Coast. Oakland, CA: Federal Emergency Management Agency, 2005:8.

64. Huang W, Xu S, Nnaji S. Evaluation of GEV model for frequency analysis of annual maximum water levels in the coast of United States. Ocean Engineering, 2008; 35(11–12):1132–1147.

65. Coles S. An Introduction to Statistical Modeling of Extreme Values. Springer Science & Business Media, 2001.

66. Coles S, Tawn J. Bayesian modelling of extreme surges on the UK east coast. Philosophical Transactions of the Royal Society of London A: Mathematical, Physical and Engineering Sciences, 2005; 363(1831):1387–1406.

67. Ministry of Infrastructure and the Environment (Rijkswaterstaat). Water Data. 2015. Available at: https://www.rijkswaterstaat.nl/water/waterdata-en-waterberichtgeving/waterdata/index.aspx

68. Gilleland E. extRemes: Extreme Value Analysis., 2015.

69. Bayes T. An Essay towards Solving a Problem in the Doctrine of Chances. Philosophical Transactions of the Royal Society of London, 1763; 53:370–418.

70. Reis DS, Stedinger JR. Bayesian MCMC flood frequency analysis with historical information. Journal of Hydrology, 2005; 313(1):97–116.

71. Metropolis N, Rosenbluth AW, Rosenbluth MN, Teller AH, Teller E. Equation of state calculations by fast computing machines. The journal of chemical physics, 1953; 21(6):1087–1092.

72. Hastings WK. Monte Carlo sampling methods using Markov chains and their applications. Biometrika, 1970; 57(1):97–109.

73. Chen M-H, Shao Q-M. Monte Carlo estimation of Bayesian credible and HPD intervals. Journal of Computational and Graphical Statistics, 1999; 8(1):69–92.

74. Saltelli A, Ratto M, Andres T, Campolongo F, Cariboni J, Gatelli D, Saisana M, Tarantola S. Global Sensitivity Analysis: The Primer. John Wiley & Sons, 2008.

75. Saltelli A. Sensitivity Analysis for Importance Assessment. Risk Analysis, 2002; 22(3):579–590.

76. Tang Y, Reed P, Wagener T, Van Werkhoven K. Comparing sensitivity analysis methods to advance lumped watershed model identification and evaluation. Hydrology and Earth System Sciences Discussions, 2007; 11(2):793–817.

77. Box GE, Hunter JS, Hunter WG. Statistics for Experimenters: Design, Innovation, and Discovery. Wiley-Interscience New York, 2005.

78. Butler MP, Reed PM, Fisher-Vanden K, Keller K, Wagener T. Identifying parametric controls and dependencies in integrated assessment models using global sensitivity analysis. Environmental Modelling & Software, 2014; 59:10–29.

79. Ma F, Zhang H, Bockstedte A, Foliente GC, Paevere P. Parameter Analysis of the Differential Model of Hysteresis. Journal of Applied Mechanics, 2004; 71(3):342–349.

80. Homma T, Saltelli A. Importance measures in global sensitivity analysis of nonlinear models. Reliability Engineering & System Safety, 1996; 52(1):1–17.

81. Von Neumann J, Morgenstern O. Theory of games and economic behavior. Bull. Amer. Math. Soc, 1945; 51(7):498–504.

82. Deering MK, Dunn CN. Implementation of Engineer Circular (EC) 1110-2-6067—USACE Process for the National Flood Insurance Program (NFIP) Levee System Evaluation. Pp. 2305–2315 in World Environmental and Water Resources Congress 2011. American Society of Civil Engineers, 2011.

83. USACE. Process for the National Flood Insurance Program (NFIP) Levee System Evaluation. Engineer Circular (EC) 1110-2-6067. Washington, D.C.: U.S. Army Corps of Engineers, 2010:11.

84. Coello CAC, Lamont GB, Van Veldhuizen DA, others. Evolutionary Algorithms for Solving Multi-Objective Problems. Springer, 2007.

85. Kotz S, Nadarajah S. Extreme Value Distributions: Theory and Applications. World Scientific, 2000.

86. Weitzman ML. Why the Far-Distant Future Should Be Discounted at Its Lowest Possible Rate. Journal of Environmental Economics and Management, 1998; 36(3):201–208.

87. Stern NH. Stern Review: The Economics of Climate Change. Cambridge University Press, 2006.

88. Nordhaus WD. A review of the "Stern review on the economics of climate change." Journal of economic literature, 2007:686–702.

89. Tuana N. Leading with ethics, aiming for policy: New opportunities for philosophy of science. Synthese, 2010; 177(3):471–492.

90. Newell RG, Pizer WA. Uncertain discount rates in climate policy analysis. Energy Policy, 2004; 32(4):519–529.

91. Louisiana Coastal Protection and Restoration Authority. Louisiana's Comprehensive Master Plan for a Sustainable Coast. Baton Rouge, 2017. Available at: http://coastal.la.gov/wp-content/uploads/2017/04/2017-Coastal-Master-Plan_Web-Single-Page_Final_Compressed-04242017.pdf

92. Hadka D, Reed P. Borg: An Auto-Adaptive Many-Objective Evolutionary Computing Framework. Evolutionary Computation, 2012; 21(2):231–259.

93. Permanent Service for Mean Sea Level. Data and Station Information for Delfzijl. Available at: http://www.psmsl.org/data/obtaining/stations/24.php

**TABLES AND FIGURES**

**Table I.** Parameter defaults, symbols, units, values, and uncertainty values used in this study. Shaded region groups parameters into Economic, Sea level rise, and Storm surge categories.

| Category | Van Dantzig (1956) Parameter (symbol) | Prior | Unit | This Study Parameter (symbol) | Distribution | Prior mean (std. dev.) | Unit |
|---|---|---|---|---|---|---|---|
| Economic | Value of goods (V) | $2 \times 10^{10}$ | Guilders | Value of goods (V) | Normal | $2 \times 10^{10}$ ($1 \times 10^9$) | Guilders |
| Economic | Effective discount rate ($\delta'$) | 0.02 | Percent/yr | Effective discount rate ($\delta'$) | Lognormal | 0.02 (0.1) | Percent/yr |
| Economic | Cost rate of heightening (k) | $4.2 \times 10^7$ | Guilders/m | Cost rate of heightening (k) | Normal | $4.2 \times 10^7$ ($4 \times 10^6$) | Guilders/m |
| Sea level rise | Subsidence rate ($\eta$) | 0.002 | m/yr | Subsidence rate ($\eta$) | Lognormal | 0.002 (0.1) | m/yr |
| Sea level rise | Sea level rise rate ($\Phi$) | 0.008 | m/yr | Sea level rise in 2015 (a) | Well-characterized joint probability distribution | -17.0–76.0 | mm |
| Sea level rise | | | | Sea level rise rate (b) | | -0.70–3.90 | mm/yr |
| Sea level rise | | | | Sea level rise acceleration (c) | | -0.0075–0.013 | mm/yr$^2$ |
| Sea level rise | | | | Year of abrupt sea level rise (t*) | Expert assessment | 2015–2090 | Year |
| Sea level rise | | | | Rate of abrupt sea level rise (c*) | Expert assessment | 0.000–0.035 | m/yr |
| Storm surge | Initial flood frequency ($p_0$) | 0.0038 | Unitless | Anomaly shape parameter ($\xi$) | Well-characterized Generalized Extreme Value distribution | -0.14–0.094 | Unitless |
| Storm surge | Exponential flood frequency rate ($\alpha$) | 2.6 | Unitless | Anomaly location parameter ($\mu$) | | 278–291 | Unitless |
| Storm surge | | | | Anomaly scale parameter ($\sigma$) | | 39.9–50.5 | Unitless |

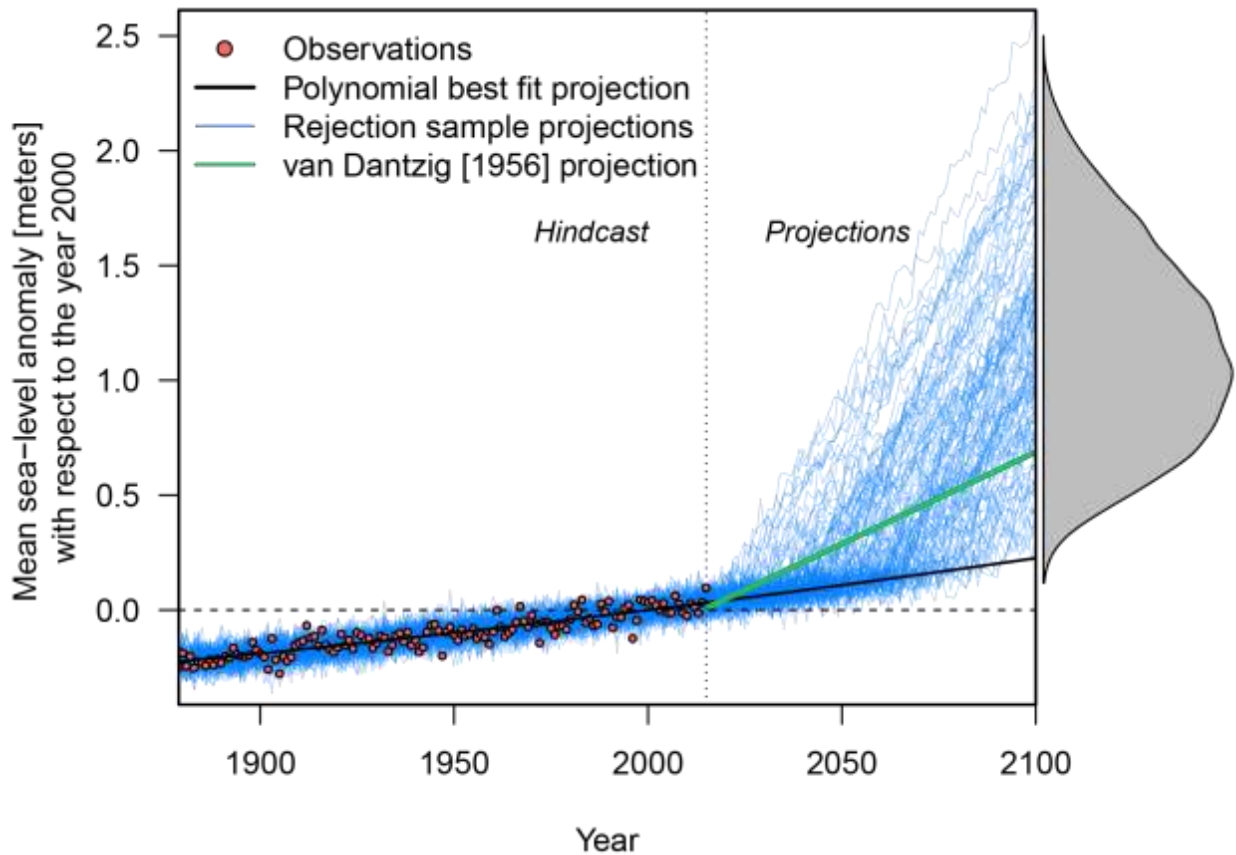

**Fig. 1.** Sea-level rise hindcasts and projections to the year 2100. Shown is the 2$^{nd}$-order polynomial best fit, calibrated by tide gauge observations (red circles). Bootstrapped residuals (blue lines) show the uncertain hindcasts and future sea-level projections. Steep increases in sea-level rate represent the potential contribution from abrupt sea-level rise. The green line shows the original van Dantzig (1956) linear sea-level approximation. The marginal beta distribution in gray approximates the expert assessment of sea-level-rise in the year 2100 (Lempert et al.[26]).

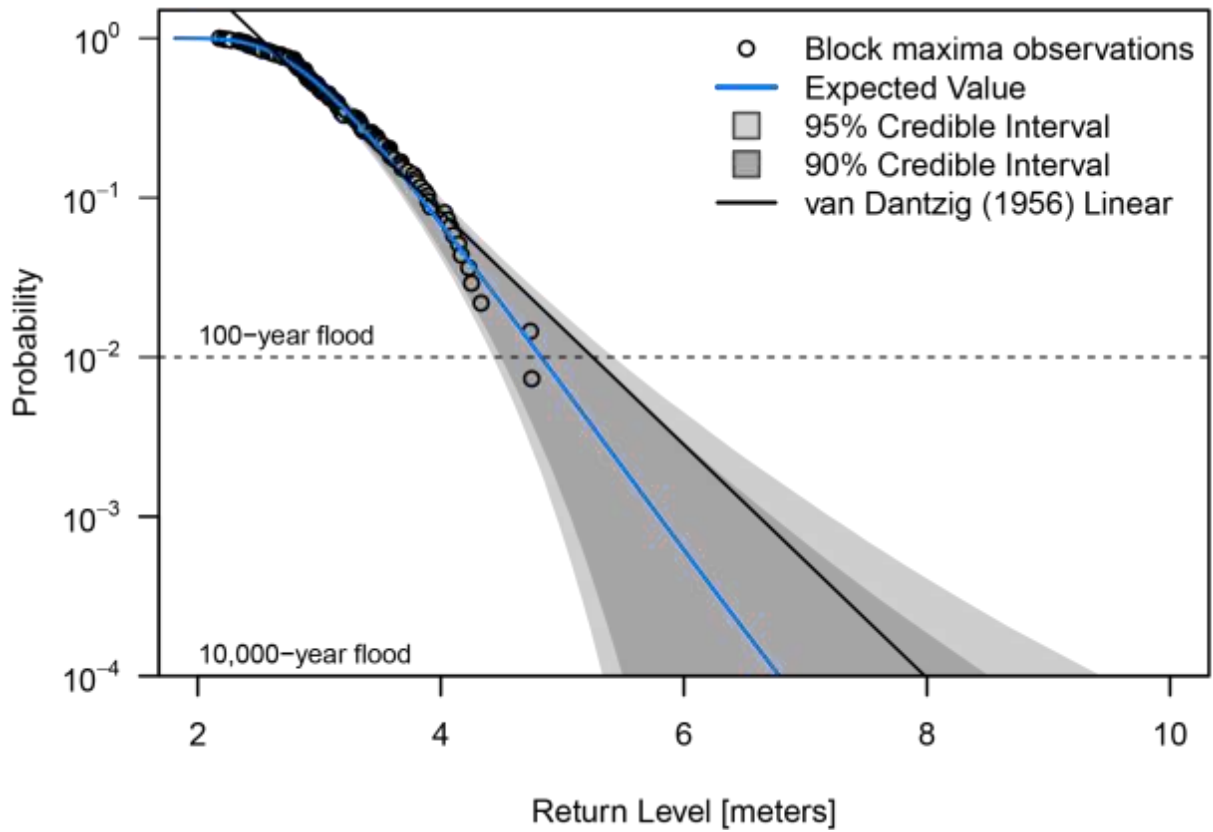

**Fig. 2.** Projected return levels using a Bayesian inversion (MCMC) technique. The circles show the annual block maxima observations from Delfzijl tide gauge.[93] Light gray envelope represents the 95% credible interval and the dark gray envelop represents the 90% credible interval. Solid black line represents the linear fit used in van Dantzig (1956), while the solid blue line represents the expected result across all states of the world.

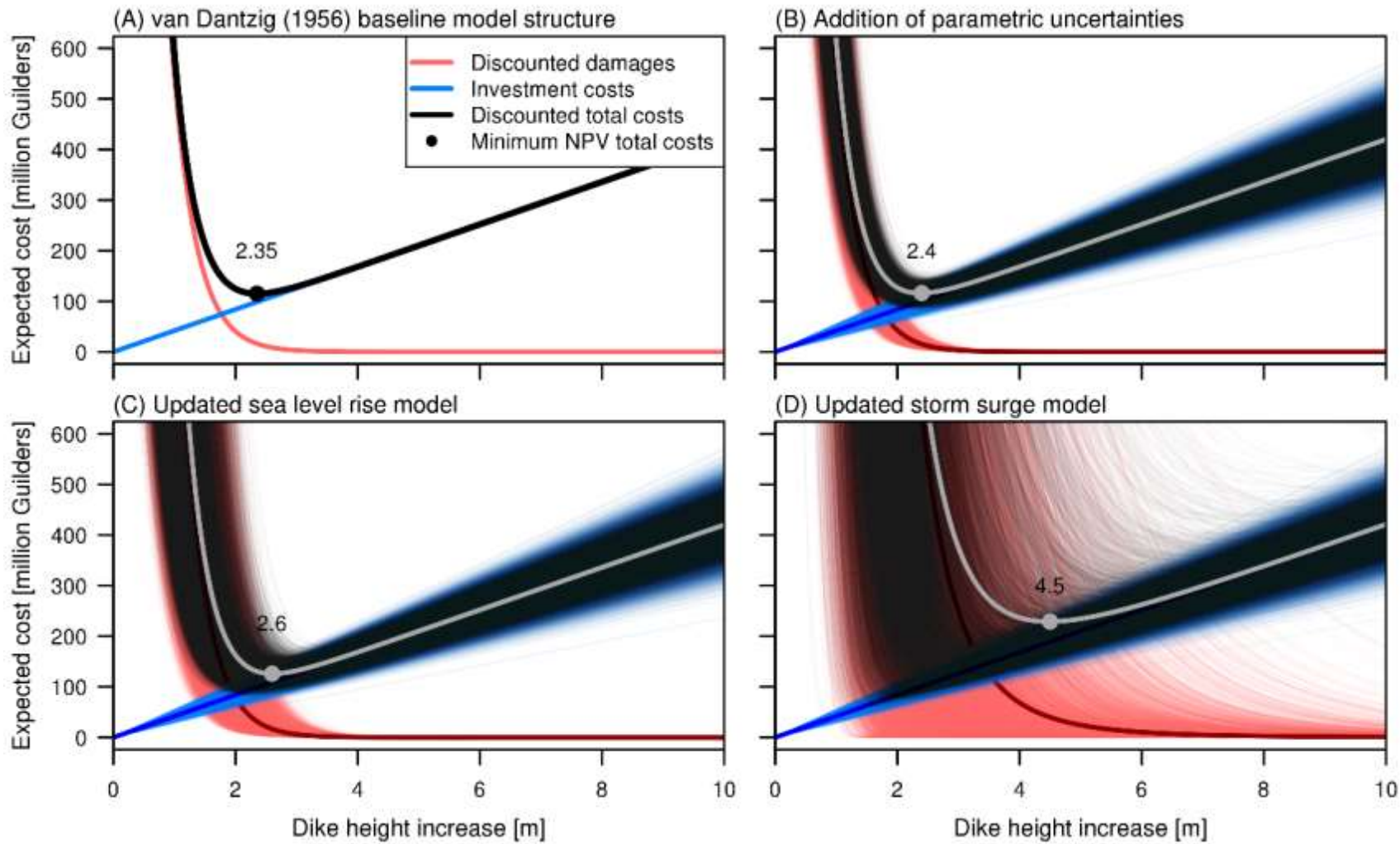

**Fig. 3.** Optimal dike heightening determined by the four considered model versions. Panel A shows baseline model version without uncertainty. The blue, red, and black curves indicate how Investment costs, Discounted damages, and Total costs, respectively, change as dike height increases. The solid point shows the optimized van Dantzig (1956) objective (to minimize the net present value of total costs). Panels B–D show the effects of incorporating increasing model uncertainties. The gray line represents expected total cost curve across 10,000 SOW, with the solid gray point showing the minimum index of the expected total costs.

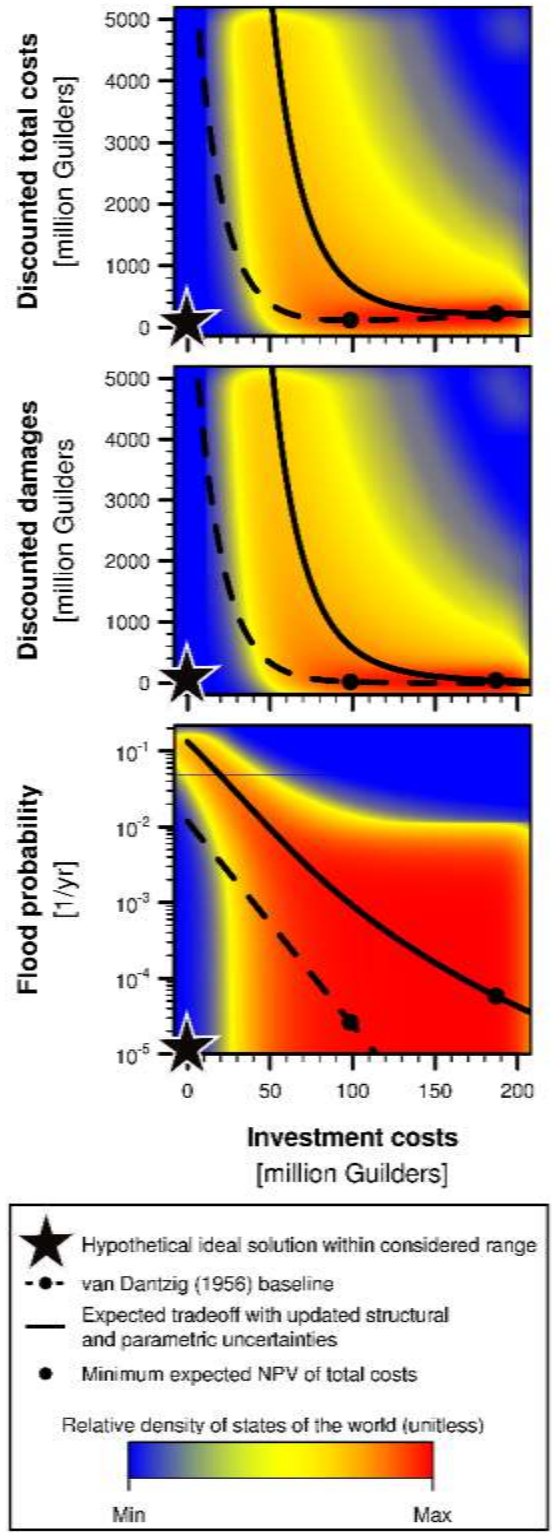

**Fig. 4.** Multi-objective trade-offs under uncertainty. Dashed lines show the baseline model trade-offs while solid lines show the expected trade-offs over 10,000 uncertain SOW. The solid points on both curves represent the minimum total costs. Color shading represents the density of determined model solutions. The black star represents the hypothetical ideal solution.

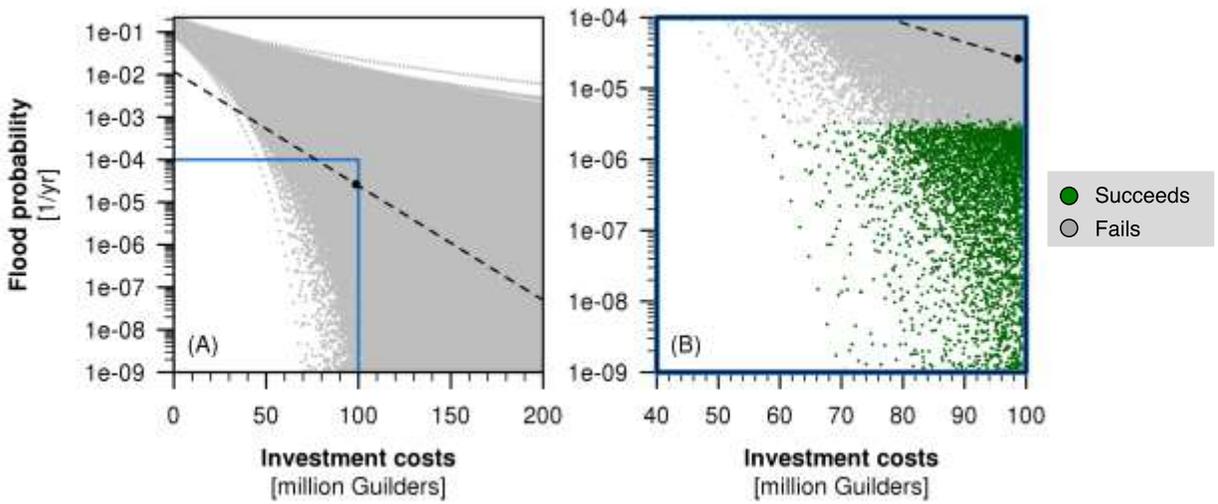

**Fig. 5.** Example of objective trade-offs using multivariate satisficing thresholds. Panel A shows solutions which satisfy the requirement of achieving flood frequencies of less than 1/10,000 per year while spending less than 100 million Guilders (blue box). Panel B shows an expanded view of that subset while instituting a third threshold (restricting flood damages to 1 million guilders – green points). The dashed line represents the van Dantzig (1956) baseline solution with the optimal solution for total costs (solid black circle).

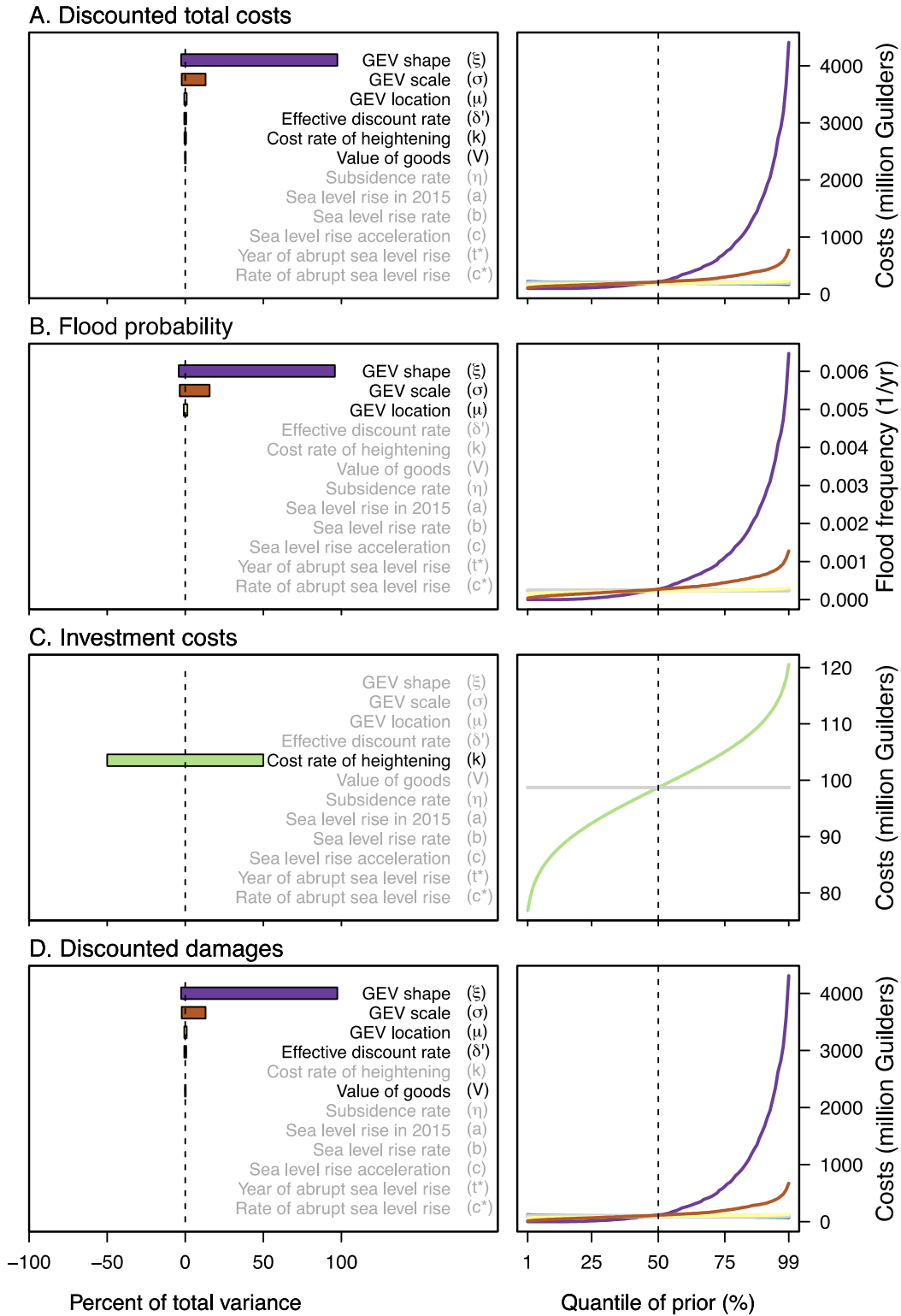

**Fig. 6.** One-at-a-time (OAT) sensitivity analysis for four management objectives with updated sea-level rise and storm surge models. Width of the bars and steepness of curve inclines indicate the degree of sensitivity to each parameter. The parameters in gray had no considerable sensitivity (less than 1% of maximum variance).

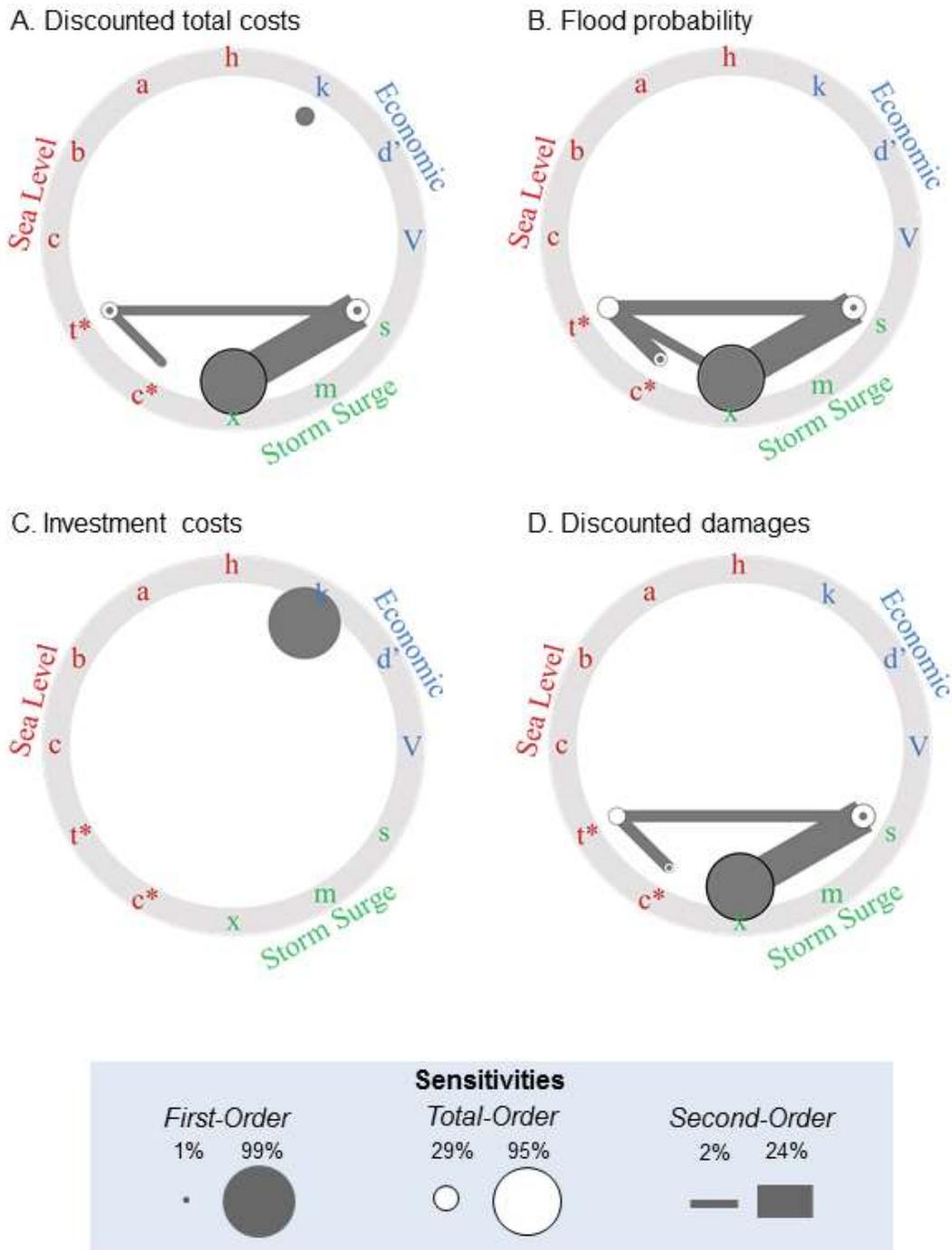

**Fig. 7.** Results of Sobol' sensitivity analysis for the four considered management objectives in the most improved model (i.e. upgraded storm surge and sea-level rise models and parameter uncertainties included). The solid circles represent the model sensitivity that can be directly attributed to a given parameter (First-Order) while connecting lines represent interactions between parameters (Second-Order). The white circles indicate Total-Order sensitivities. The sizes of the circles at each node and the widths of the lines indicate the magnitude of the sensitivities.

**SUPPLEMENTAL FIGURES**

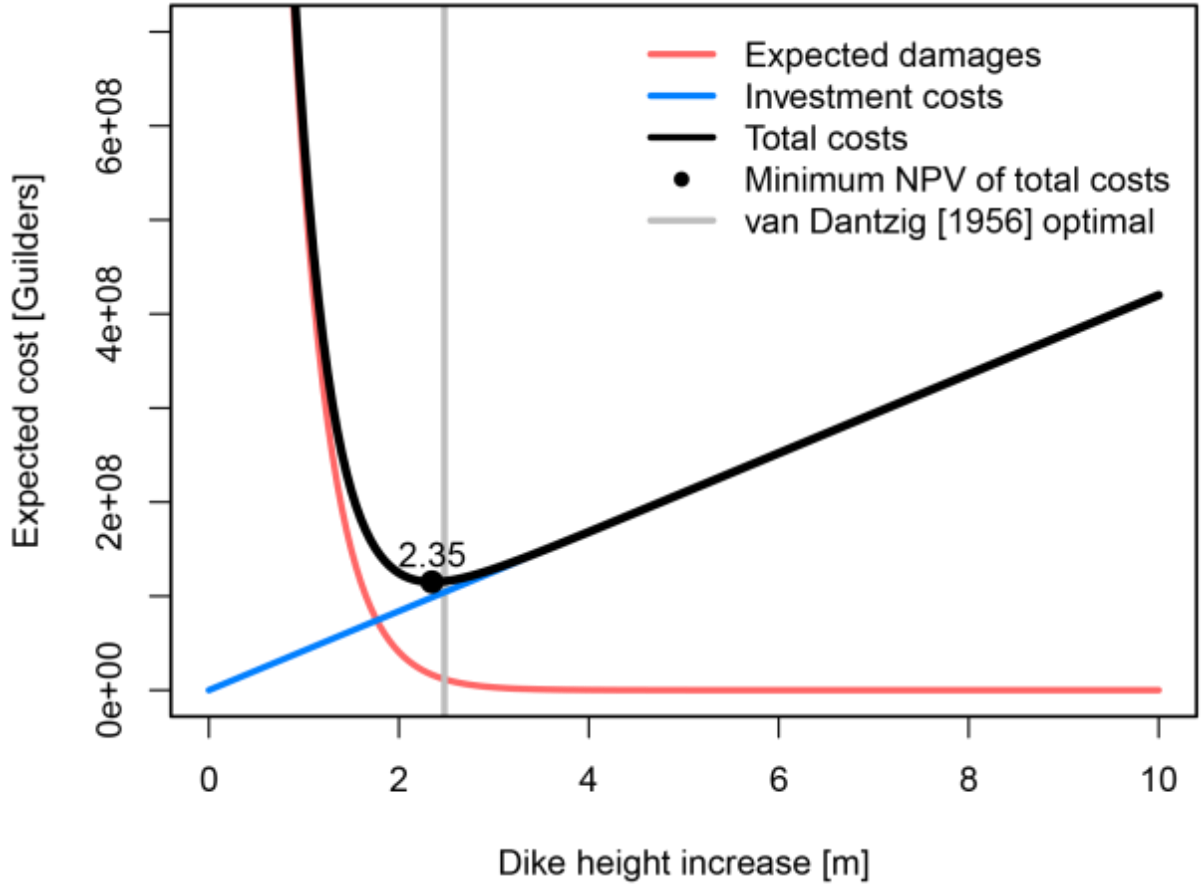

**Fig. S1.** Baseline van Dantzig model, evaluated for dike heights increases from zero to ten meters in five-centimeter increments. Vertical gray bar indicates the optimal heightening of the original publication (2.48 meters, compared to the optimal result (2.35 meters) in this implementation).

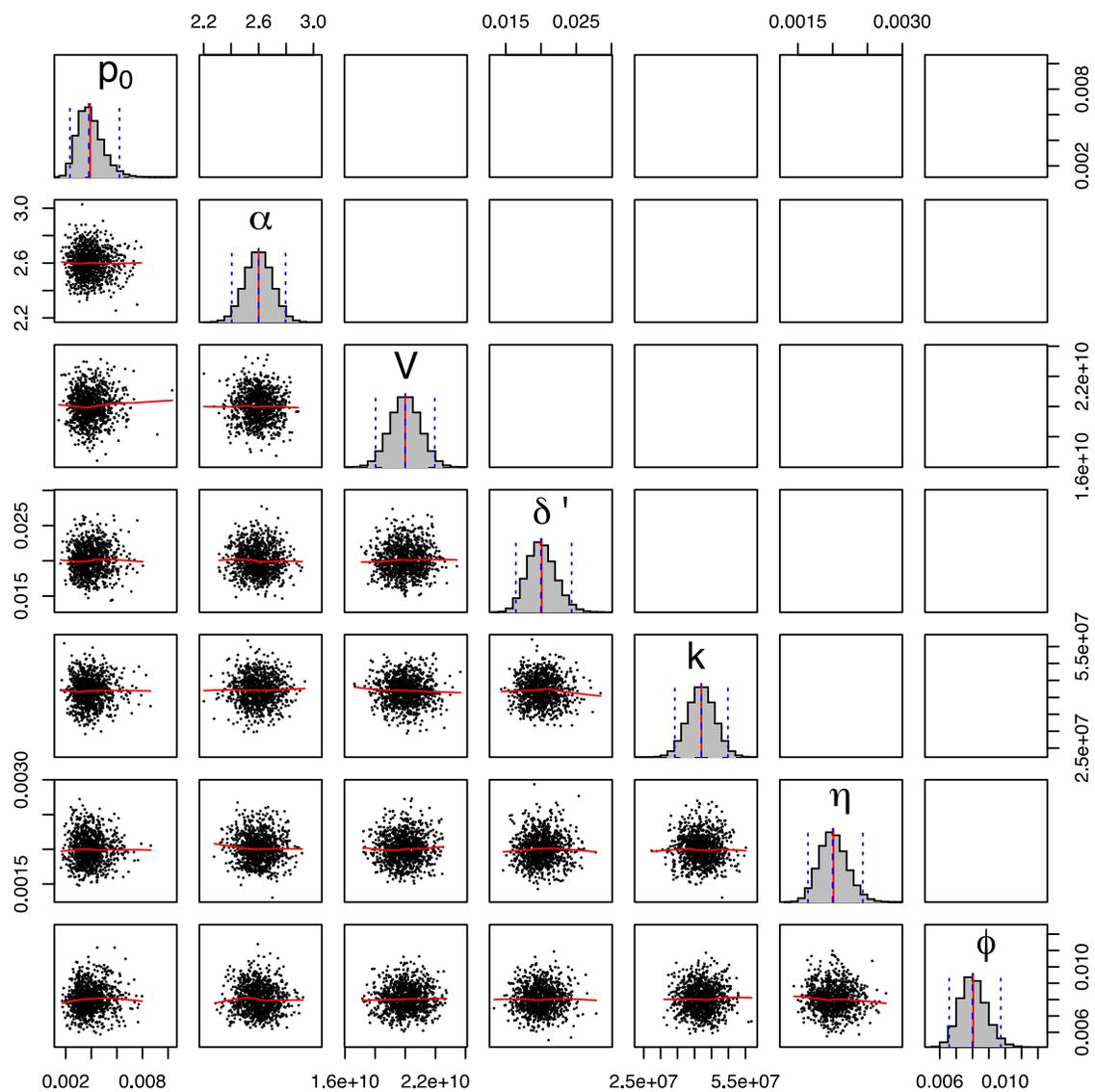

**Fig. S2.** Pairs plots and marginal distributions for the parameters used in the parametric uncertainty model. The red line shows a loess regression through the points.

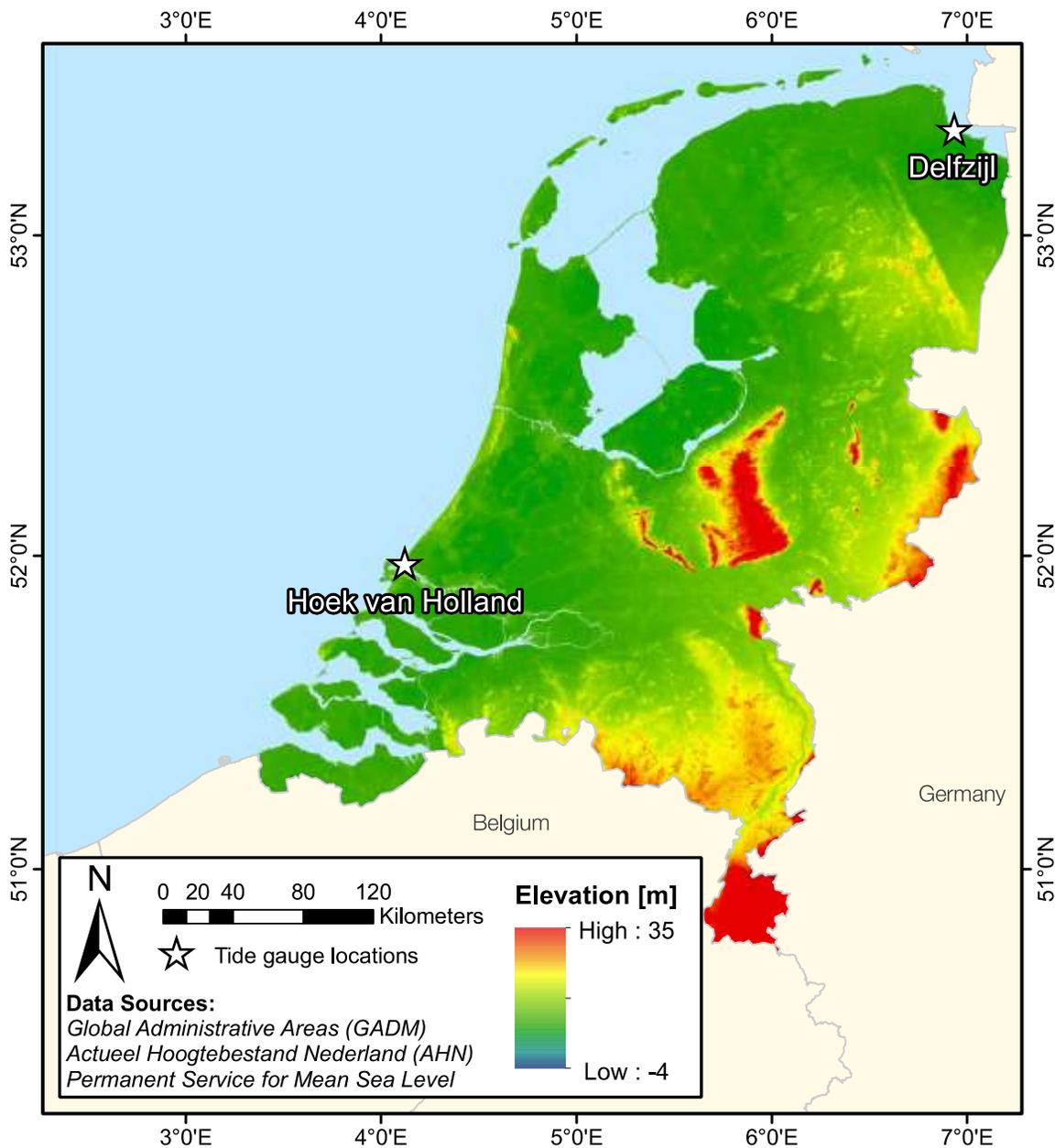

**Fig. S3.** Locations of Hoek van Holland and Delfzijl tide gauges used in the van Dantzig (1956) and this study, respectively. Tide gauge locations taken from the Permanent Service for Mean Sea Level (PSMSL). Digital elevation data taken from the Up-to-date Height Model of The Netherlands (Actueel Hoogtebestand Nederland - AHN).

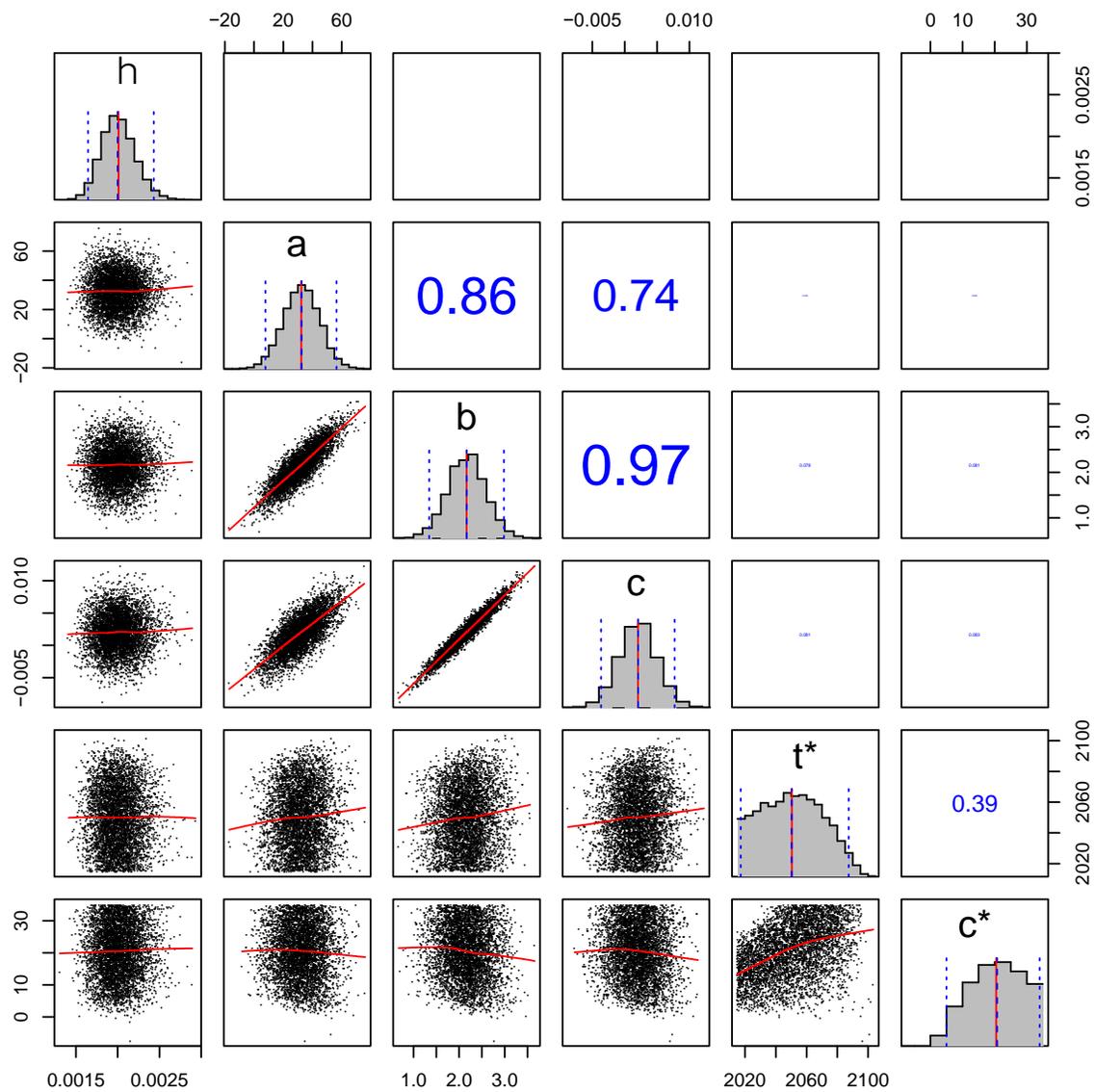

**Fig. S4.** Pairs plots and marginal distributions for the parameters used in the sea-level rise model. Blue numbers indicate correlation coefficients between parameters. The red line shows a loess regression through the points.

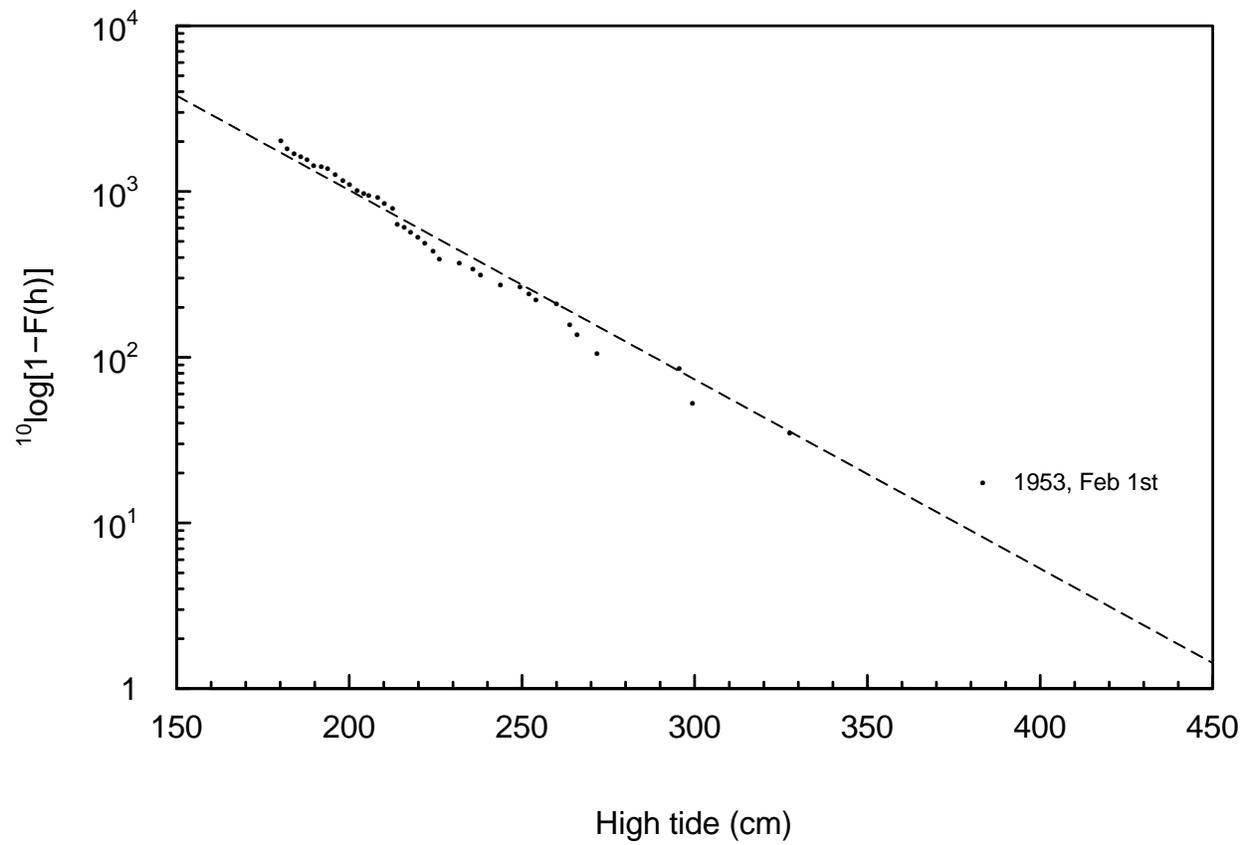

**Fig. S5.** Exceedance probabilities of annual high tides at Hoek van Holland, Netherlands from 1888-1937. Digitized from data in Fig. 1, in van Dantzig (1956).

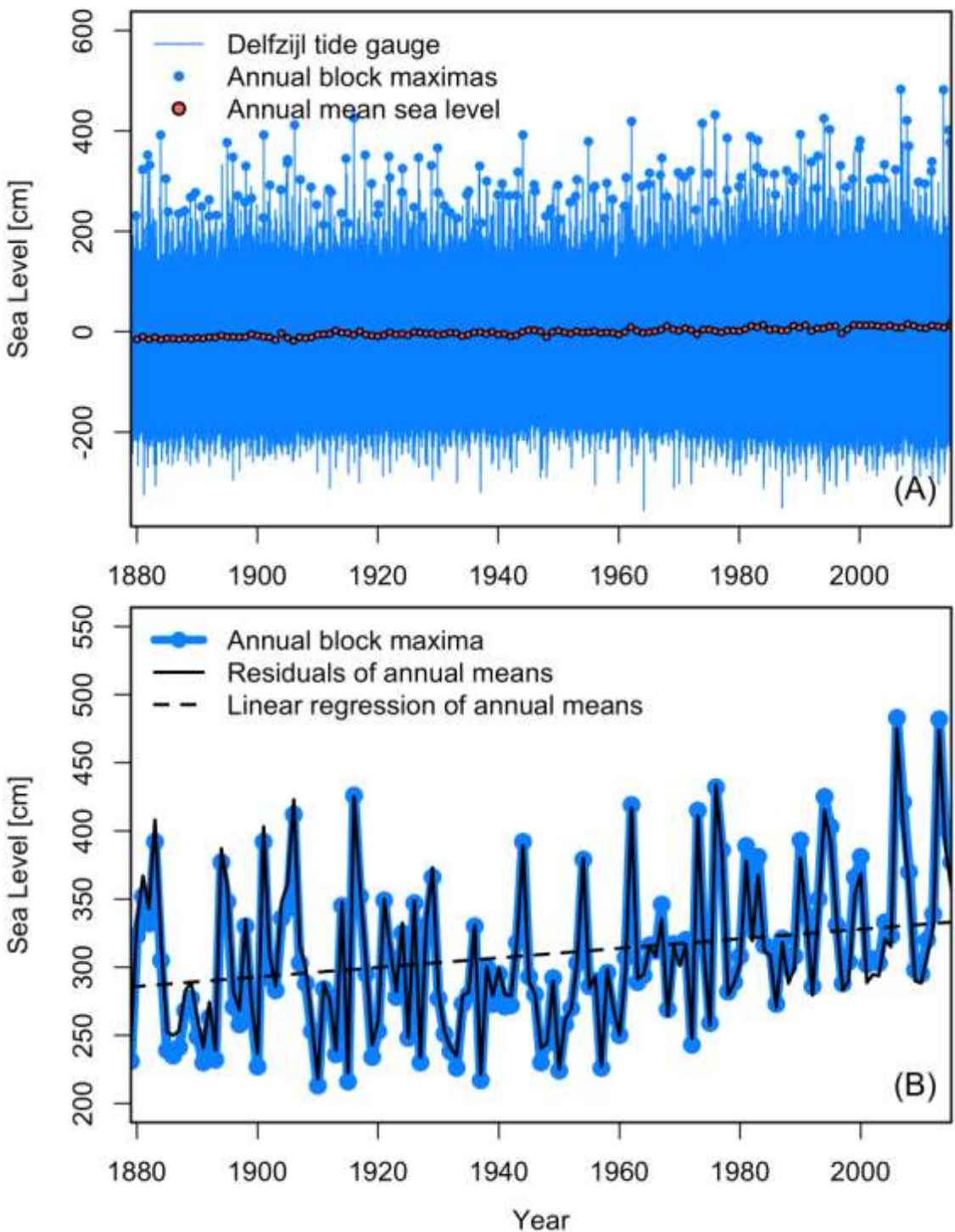

**Fig. S6.** Tide gauge observations from Delfzijl, the Netherlands. Panel A shows the raw data with the annual block maxima (solid blue circles) and annual means (filled red circles). Panel B enlarges the annual block maxima (raw observations in blue) and shows the detrended maxima after taking the residuals of the annual means (solid black lines). The dashed black line represents the linear regression through the detrended block maxima.

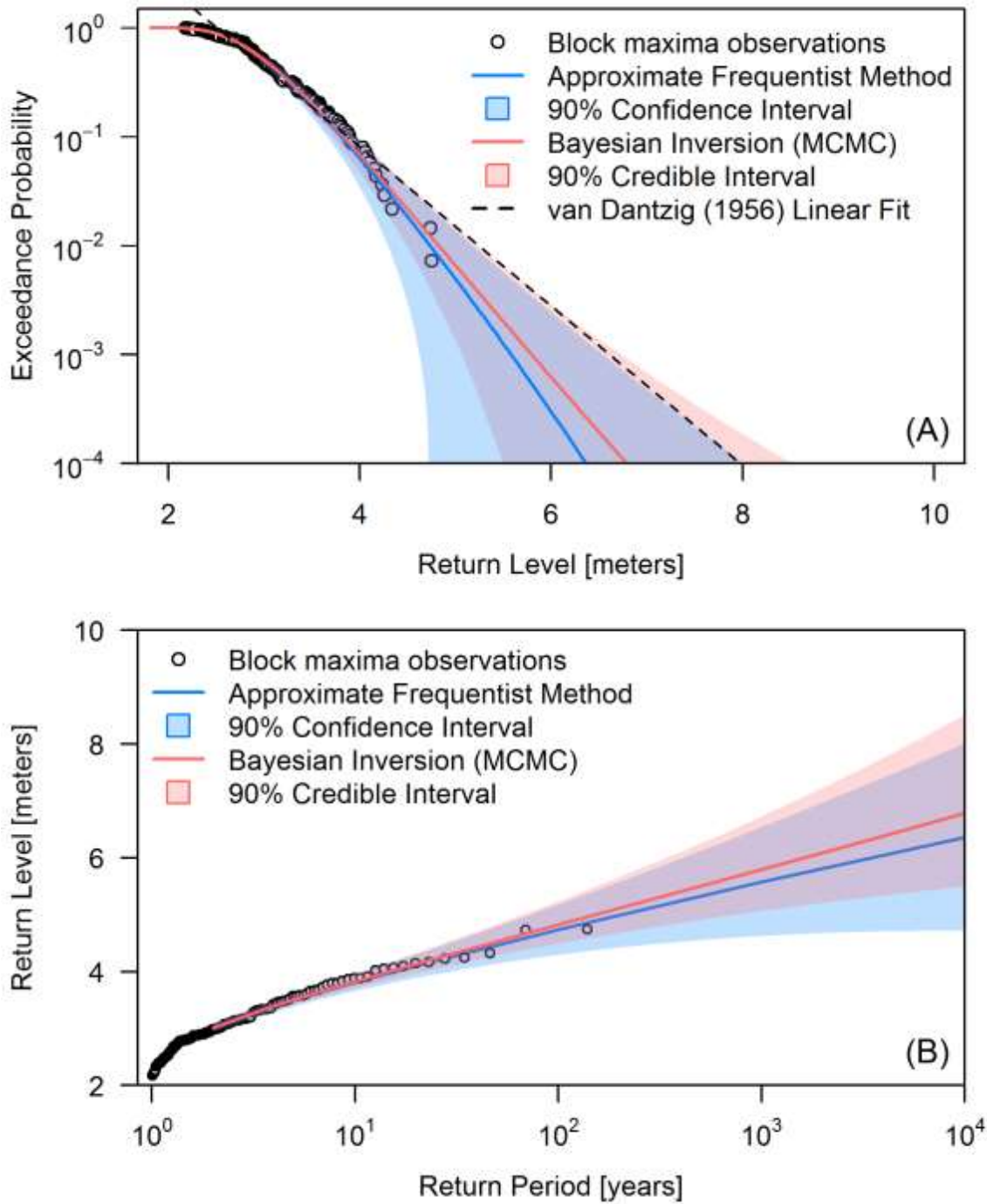

**Fig. S7.** Observed and projected return levels (Panel A) and return periods (Panel B) for Delfizjl tide gauge, the Netherlands. The shaded envelopes represent the 90% confidence interval and the 90% credible interval for the maximum likelihood estimation and Markov Chain Monte Carlo methods, respectively. Solid lines represent expected outcome for each method. Dashed line represents the linear extrapolation used in van Dantzig (1956).

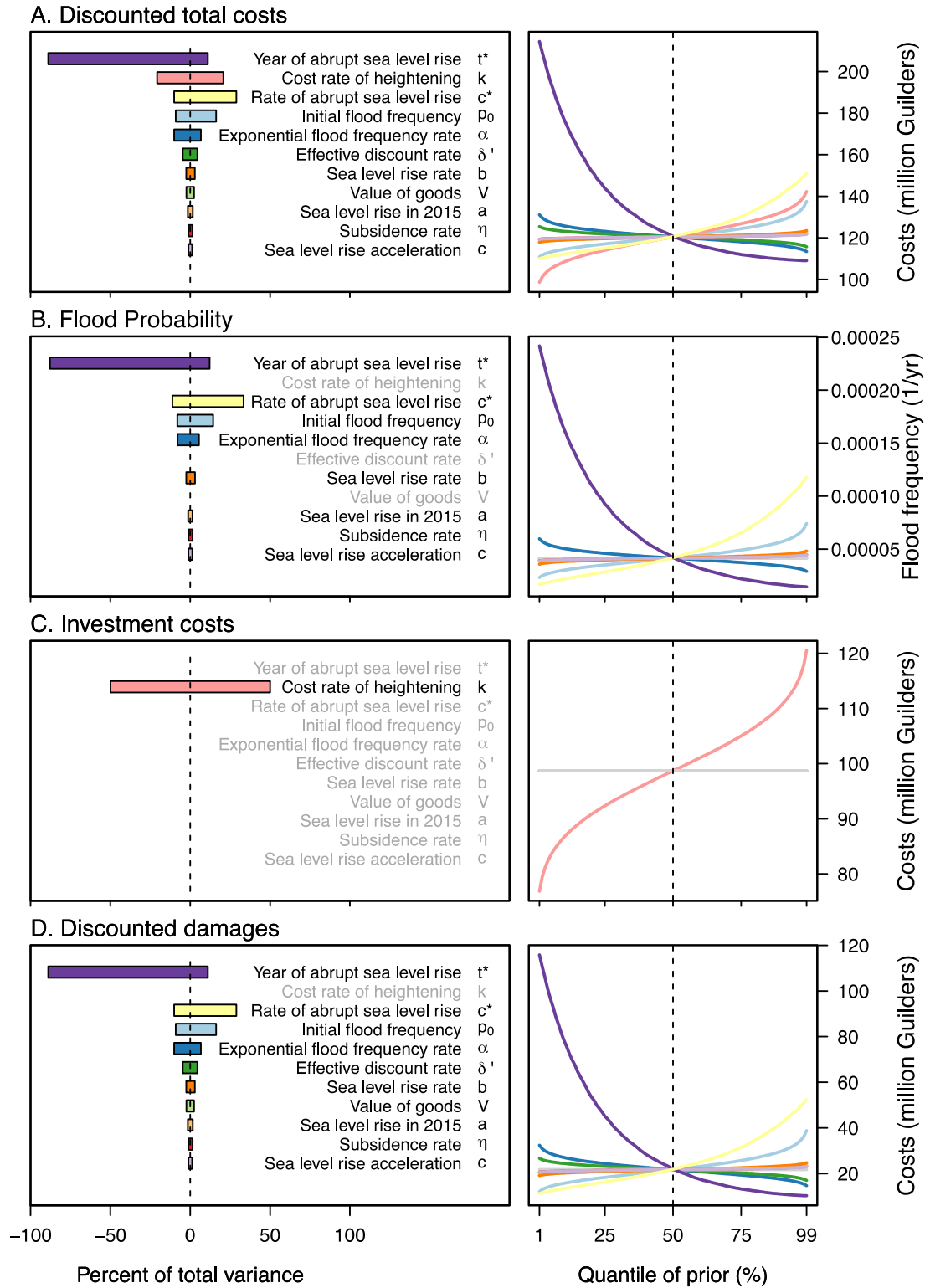

**Fig. S8.** One-at-a-time (OAT) sensitivity analysis for four management objectives with upgraded sea-level rise model. Width of the bars and steepness of curve inclines indicate the degree of sensitivity to each parameter. The parameters in gray had no considerable sensitivity (less than 1% of maximum variance).

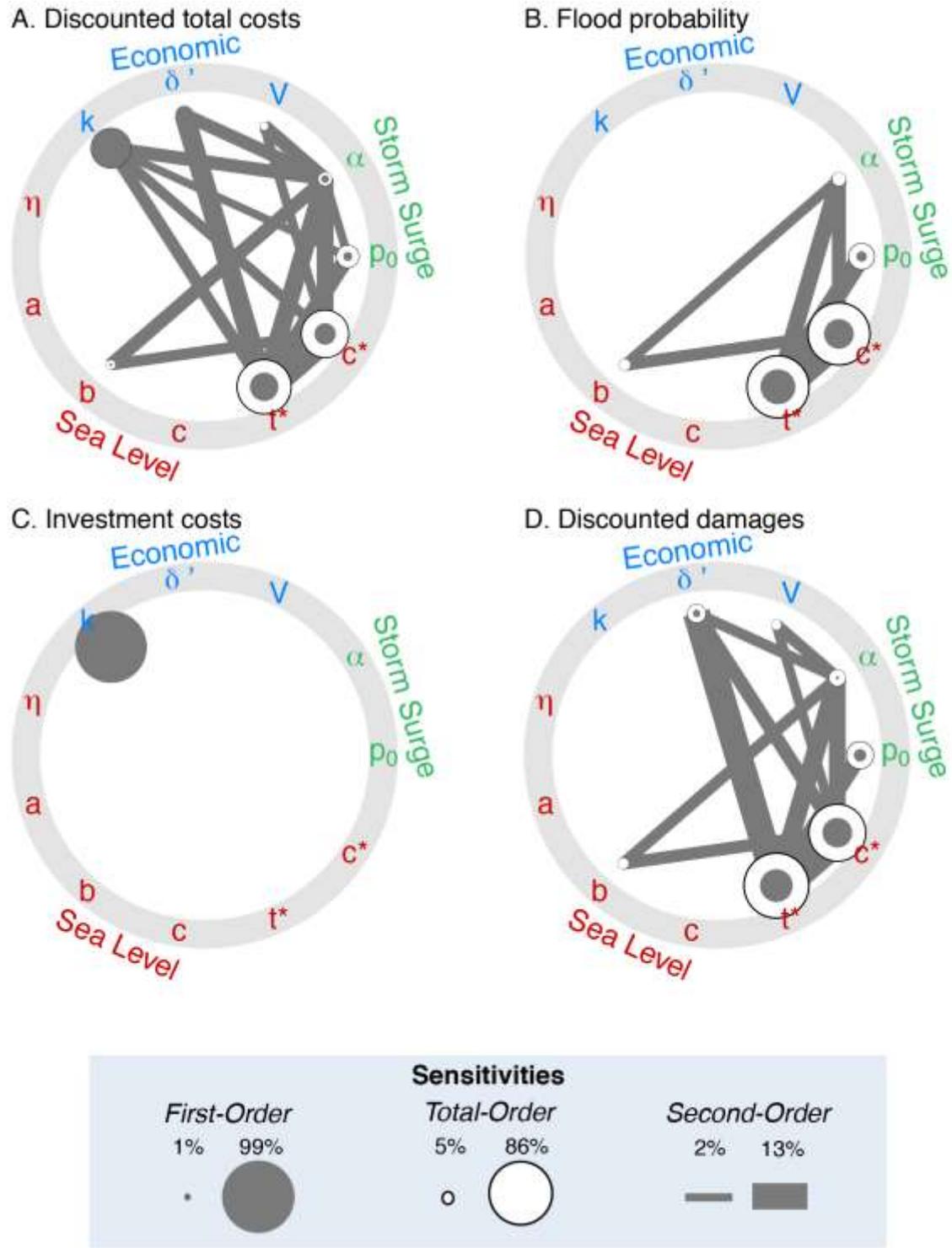

**Fig. S9.** Results of Sobol' sensitivity analysis for the four considered management objectives in the upgraded sea-level rise model. The solid circles represent the model sensitivity that can be directly attributed to a given parameter while connecting lines represent interactions between parameters. The white circles indicate total order sensitivities. The sizes of the circles at each node and the widths of the lines indicate the magnitude of the sensitivities.

46